\documentclass[10pt]{iopart}
\usepackage{iopams}  
\usepackage{amssymb}
\usepackage[T1]{fontenc}
\usepackage{setstack}
\usepackage{graphicx}
\usepackage{bm}
\usepackage{color}
\usepackage{hyperref}
\usepackage{braket}
\usepackage{comment}
\usepackage{inputenc}
\usepackage{comment}
\usepackage{bbm}
\usepackage{cite}
\begin{document}
	
	\title[Flat bands without twists: periodic holey graphene]{Flat bands without twists: periodic holey graphene}
	
	\author{Abdiel de Jesús Espinosa-Champo$^{1,2}$ and Gerardo G. Naumis$^{2}$}
	
	\address{${}^{1}$Posgrado en Ciencias Físicas, Instituto de F\'isica, Universidad Nacional Aut\'onoma de M\'exico (UNAM). Apdo. Postal 20-364, 01000, CDMX, M\'exico.}
	\address{${}^{2}$Depto. de Sistemas Complejos, Instituto de F\'isica, Universidad Nacional Aut\'onoma de M\'exico (UNAM). Apdo. Postal 20-364, 01000, CDMX, M\'exico.}

	\ead{naumis@fisica.unam.mx}
	\vspace{10pt}
	\begin{indented}
		\item[]\today
	\end{indented}
	
	\begin{abstract}
\textit{Holey Graphene} (HG) is a widely used graphene material for the synthesis of high-purity and highly crystalline materials. In this work, we explore the electronic properties of a periodic distribution of lattice holes, demonstrating the emergence of flat bands with compact localized states. It is shown that the holes  break the bipartite sublattice and inversion symmetries, inducing gaps and a nonzero Berry curvature.  Moreover, the folding of the Dirac cones from the hexagonal Brillouin zone (BZ) to the holey superlattice rectangular BZ of HG with sizes proportional to an integer $n$ times the graphene's lattice parameter leads to a periodicity in the gap formation  such that  $n \equiv 0$ (mod $3$). Meanwhile, it is shown that if $n \equiv \pm 1$ (mod $3$), a gap emerges where Dirac points are folded along the $\Gamma-X$ path. The low-energy hamiltonian for the three central bands is also obtained, revealing that the system behaves as an effective  $\alpha-\mathcal{T}_{3}$ graphene material. Therefore, a simple protocol is presented here that allows obtaining flat bands at will. Such bands are known to increase electron-electron correlated effects. This work provides an alternative system, much easier to build than twisted systems, to obtain highly correlated quantum phases. 
\end{abstract}

	\maketitle

	\section{Introduction. \label{Sec:Introduction}}
	
	In recent years, the concept of electronic flat bands has gained prominence in the fields of materials science and condensed matter physics \cite{QiuWenXuan,Drost_Ojanen_Harju_Liljeroth_2017, Abilio1999, HidekiOzawa2015,Nakata2012, He_Mao_Cai_Zhang_Li_Yuan_Zhu_Wang_2021, Cao2018, Leonardo2021, Leonardo2022, Espinosa-Champo_2024}. This heightened interest can be attributed to the electronic, geometrical and topological properties inherent to flat bands \cite{Bergholtz_Liu_2013,Nguyen2018,DENG2003412,Espinosa-Champo_2024} and their potential to give rise to novel phenomena \cite{Mielke_Tasaki_1993, Tasaki_1998, Cao2018,AokiHideo2020, WuCongjun2007,Jaworowski_2018, Leonardo2023, Leonardo2023RevMex}. Specifically, the discovery of superconductivity in twisted bilayer graphene (TBLG) in 2018 fueled considerable attention in the field \cite{Cao2018}.
	However, the primary challenge lies in the experimental control of the rotation angle. For this reason, alternative systems are currently being explored to preserve flat bands, without relying solely on the rotation angle. Examples of these systems include graphene subjected to mechanical deformations \cite{RomanTaboada2017,RomanTaboada2017b,RomanTaboada_2017JPC,Mao2020, Manesco_2021, Manesco_2021_2, Milanovic2020, Sandler2023, Elias2023} , multilayer graphene twisted (MTLG) under pressure \cite{Guinea2017, Carr2018, Yndurain2019, Wu2021Pressure, Francisco2023}, cyclicgraphyne,  cyclicgraphdiyne \cite{You2019} and holey graphene \cite{Sedelnikova2019, Mahmood2015,Zhao2017, Omidvar2017}, where the theoretical possibility of chiral superconductivity emergence has been demonstrated \cite{Sousa2022}.

	Holey Graphene (HG), also called Graphene Nanomesh (GNM), refer to the presence of holes distributed in the material \cite{Yang2023, Mahmood2015, Lin2023, Liu2022}. Introducing holes in graphene provides an additional degree of freedom to control optoelectronic and mechanical properties\cite{Naumis2007,Xu2019,Singh2020}.
	For example, holes create a mobility gap in graphene making it a narrow gap semiconductor \cite{Naumis2007}. This opens opportunities for exploring electronic, optical and correlated dependent phenomena and designing graphene-based devices that can exploit the unique properties of holey graphene \cite{Rapjut2023, Lokhande2023, Rivera2021, Liu2020, YiLin2017}. 
	
	Furthermore, the size and distribution of the holes can also modify the electronic properties. Using DFT calculations, Barkov et al. \cite{PavelBarkov2021} investigated holey graphene's transmission function, $T(E)$. They found that introducing periodic holes can modulate the values of $T(E)$ by changing the distances between holes and maintaining the hole size; thus, the conductance tends to increase farther apart. 
	
	In addition, several techniques are employed to fabricate HG, including electron/ion beam lithography \cite{Lokhande2023, Fischbein2008, Rapjut2023}, plasma etching \cite{Lokhande2023,Rapjut2023}, chemical etching \cite{Mahmood2015}, laser-based methods \cite{Khan2021, KAZEMIZADEH2018, Wang2019, Lin2014, KUMAR2022, Joshi2022}, and others \cite{Liu2022, HaruyamaJunji2013, Lokhande2023, Rapjut2023, LinYi2022,YiLin2017, Rivera2021, Zhang2019, White2020, Zhang2016} . These techniques offer an experimental advantage over  MTLG. Among them, the most common techniques used to fabricate HG are:  
	
	Top-down lithography involves the selective removal of graphene sheets to create nanopores with uniform sizes and spatial arrangement. Although successful in producing graphene sheets with uniform nano-size pores, this technique faces challenges in large-scale fabrication and requires specialized equipment and expertise, making it expensive \cite{Lokhande2023, Fischbein2008, Rapjut2023}. 
	
	Another technique used to fabricate HG is template-assisted chemical vapor deposition \cite{Lokhande2023, Fischbein2008, YiLin2017, Zhang2019}. In this method, HG is grown on specific inorganic templates. This technique allows for the bottom-up growth of HG and offers control over nanopores' size, shape, and distribution \cite{Zhang2019}. However, these fabrication techniques have certain drawbacks because they require careful control of various parameters such as temperature, pressure, and duration, making the process lengthy, cumbersome, and costly.

	A third approach to synthesizing HG is using block copolymer lithography techniques on CVD-grown graphene \cite{HaruyamaJunji2013, Bai2010}.
	This technique involves patterning the graphene with a copolymer material and using reactive ion etching to create nanopores in the desired arrangement. One advantage of this technique is that it allows for the control of the neck width, which can be important in determining the properties and applications of HG. However, this approach still requires further optimization to achieve large-scale production of HG. 
	
Nevertheless, laser-based methods have recently emerged as a promising technique for synthesizing HG \cite{Khan2021, KAZEMIZADEH2018, Wang2019, Lin2014, KUMAR2022, Joshi2022}. These methods offer a precise means of controlling the synthesis process of the holey structure. Researchers can adjust parameters like laser intensity and target material to create holes with specific sizes, shapes, and distributions. This level of control enables the production of HG with tailored properties for various applications. Moreover, this method allows for the fabrication of high-purity and highly crystalline HG without the need for catalysts or chemicals. Laser ablation methods are also scalable, making them ideal for large-scale HG production. Furthermore, these techniques permit the customization of HG by introducing dopants and functional groups. This versatility contributes to the development of graphene-based devices with enhanced performance. However, it's worth noting some notable disadvantages, including the cost and complexity of the equipment, limited scalability for large-area synthesis of pristine graphene, and challenges in achieving structural uniformity.

In this work, we explore different configurations of holey graphene which we show are able to produce flat-bands. From there, we study its electronic properties of the resulting lattices.  The structure of this work is as follows: in Section \ref{sec:model-tight-binding}, we present the tight-binding model and the unit cells of the holey graphene primarily under discussion. In Section \ref{sec: electronic properties} we present the electronic properties of the systems introduced in Section \ref{sec:model-tight-binding}. Therein, the emergence of flat bands with compact localized states and energy gaps  is discussed  in terms of the Brillouin zone folding and bipartite sublattices site imbalance.  Finally, our conclusions and future perspectives are outlined in Section \ref{sec:conclusions}.

	\section{Tight-binding holey graphene model\label{sec:model-tight-binding}}
	
\begin{figure}
\fl	
  a) \includegraphics[height=0.28\textheight]{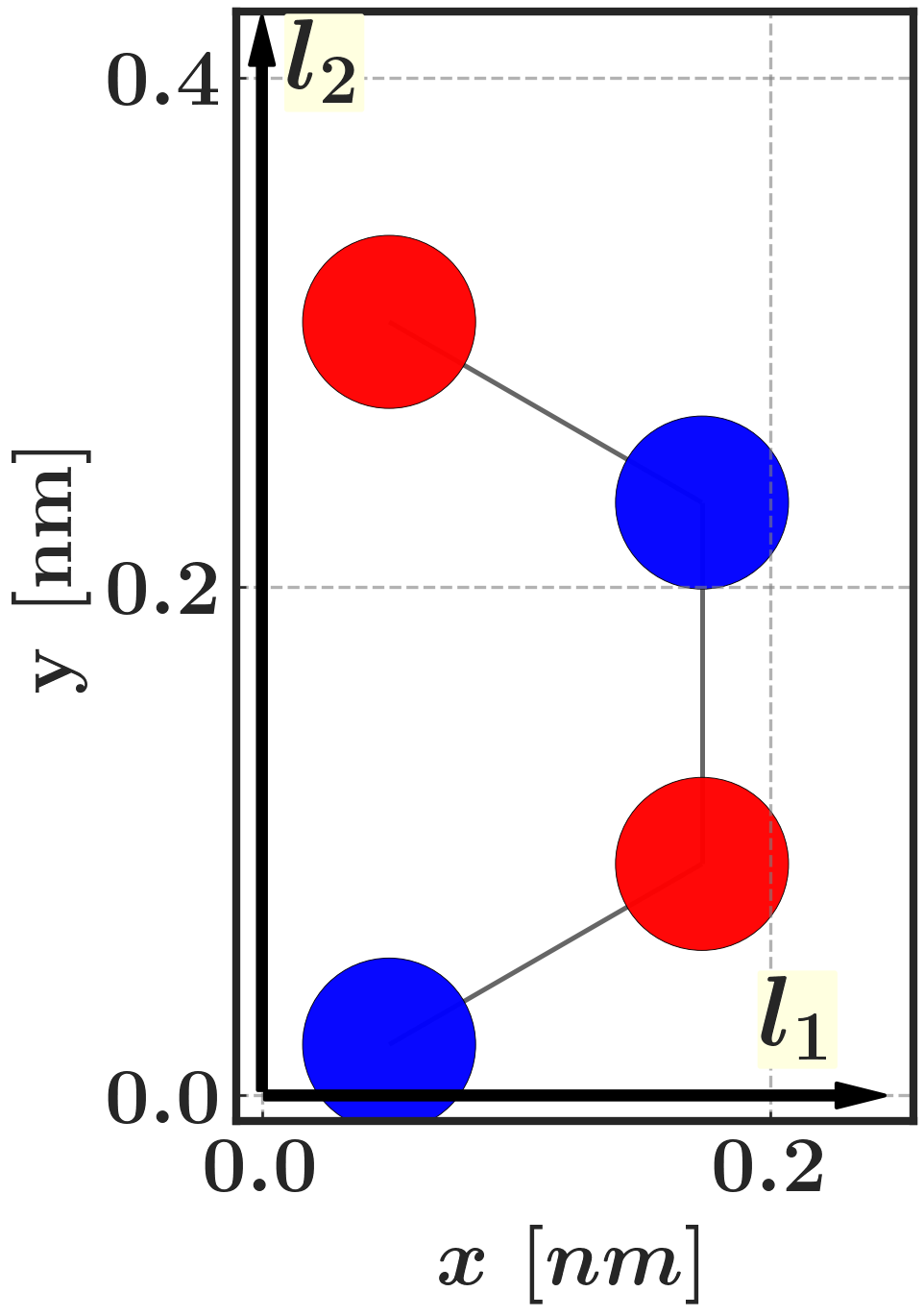}
b) \includegraphics[height=0.28\textheight]{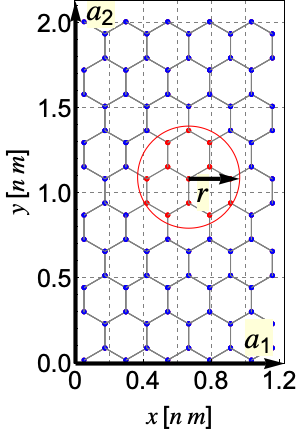}
c) \includegraphics[height=0.28\textheight]{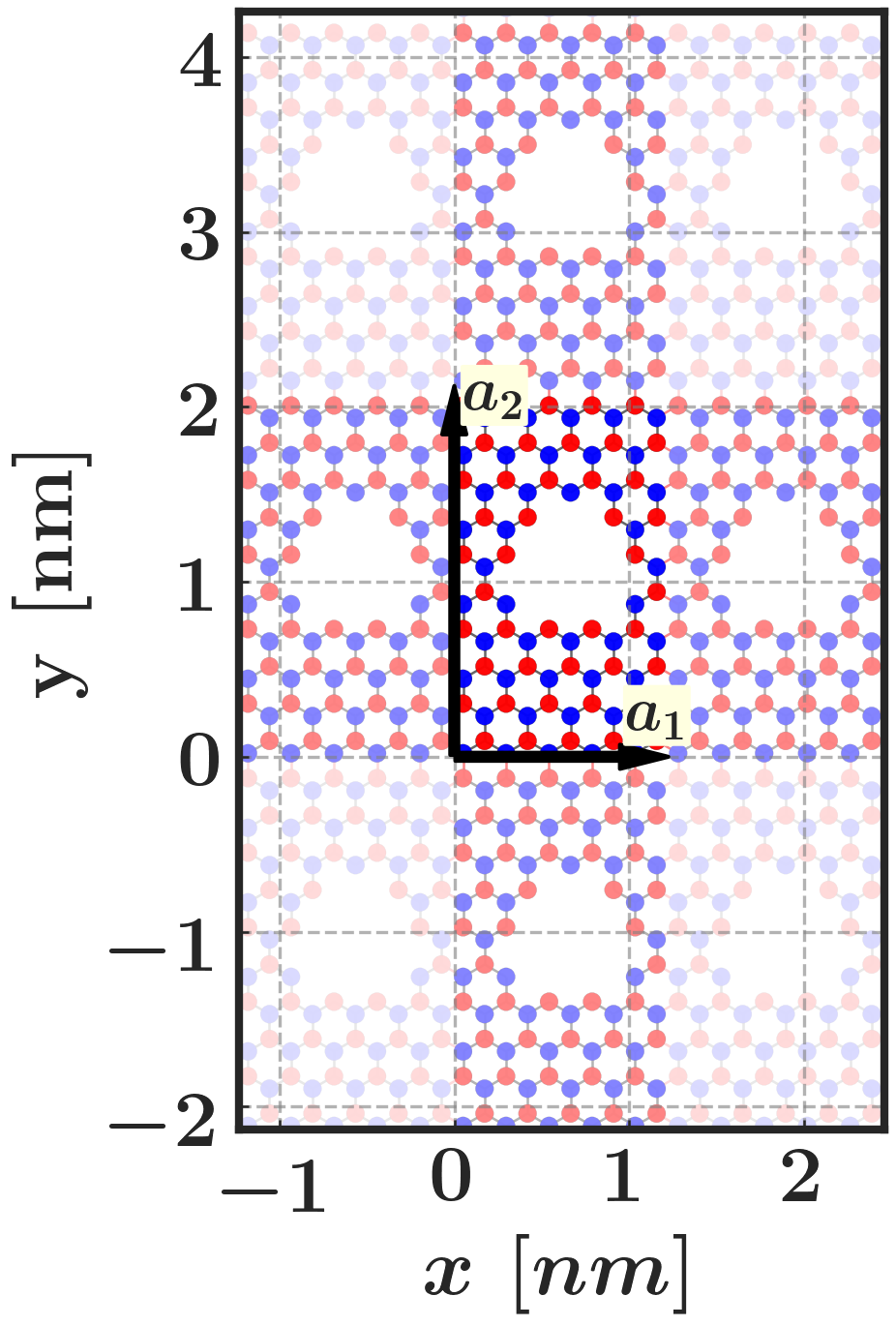}\\
\fl
d) \includegraphics[height=0.2\textheight]{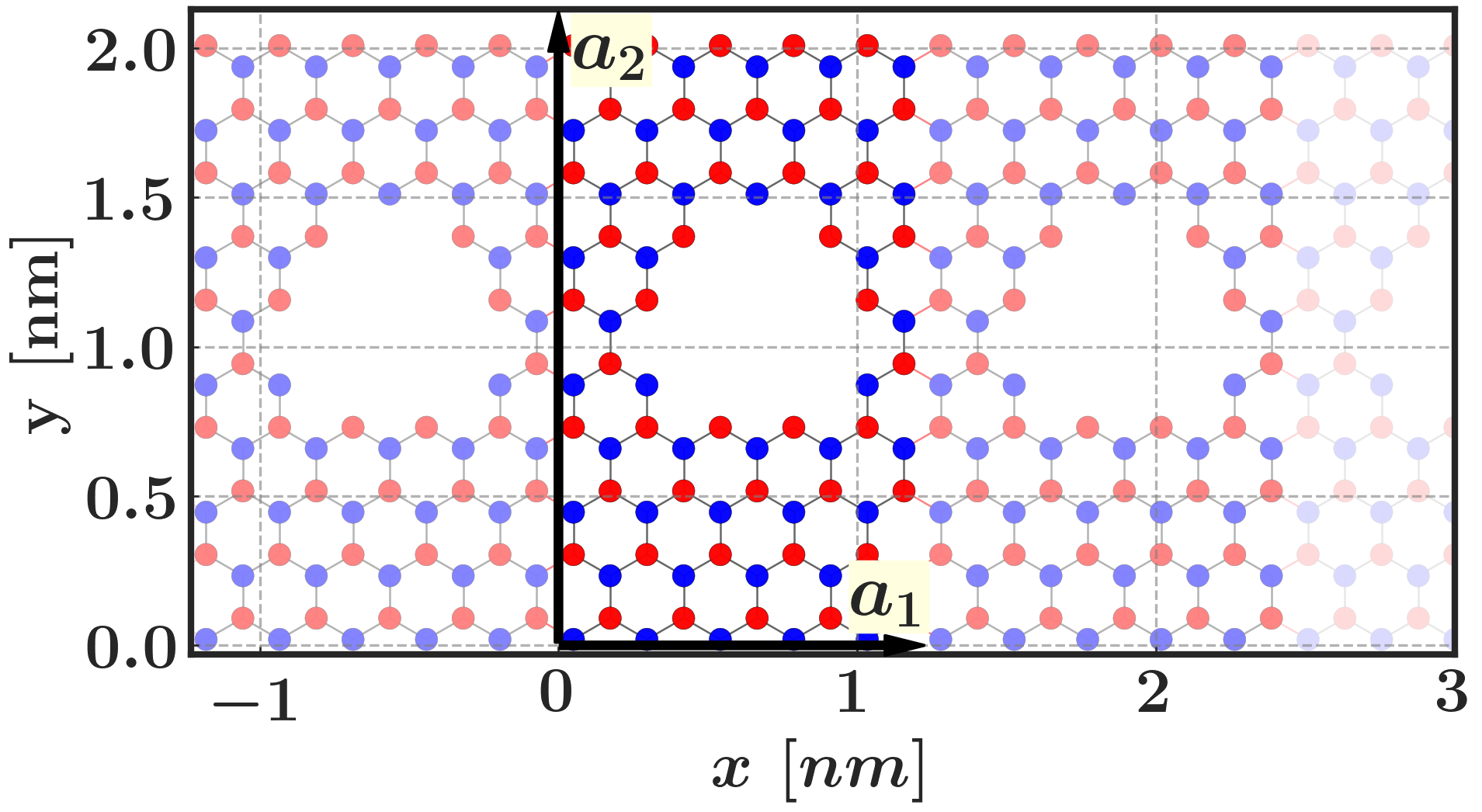}
e) \includegraphics[height=0.3\textheight]{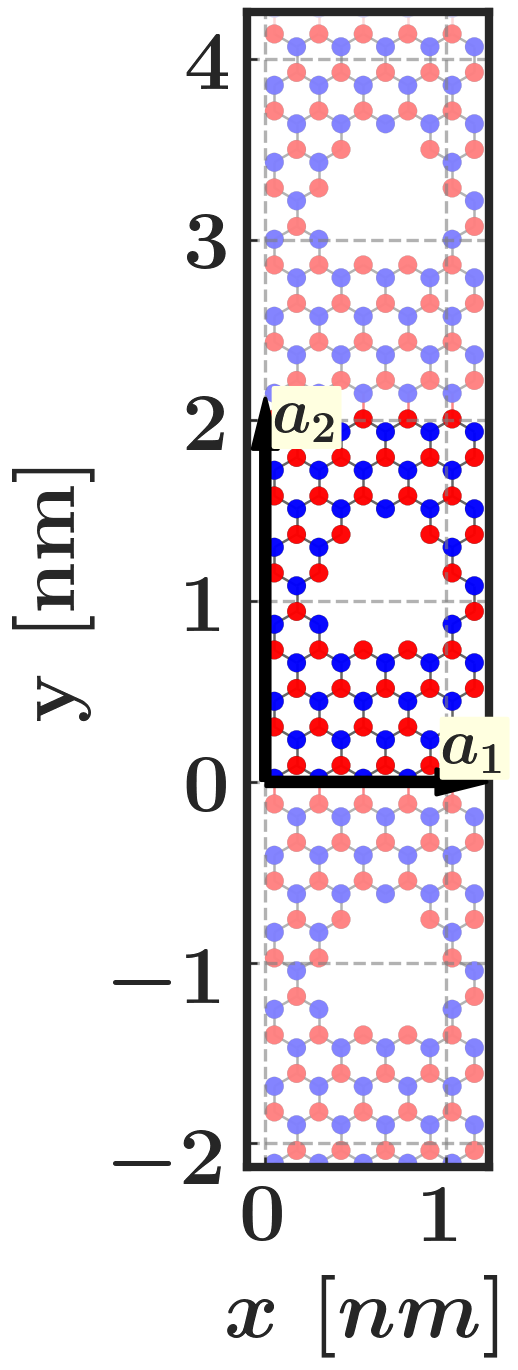}
		\caption{ a) Unit cell of graphene with a 4-atom basis and lattice vectors $\boldsymbol{l}_{1}=(\sqrt{3} a{0}),\boldsymbol{l}_{2}=(0,3a_0)$ arranged such that the zigzag and armchair directions align with the horizontal and vertical axes, respectively. b) Unit cell for hollow graphene, featuring lattice vectors $\boldsymbol{a}_{1}=n \boldsymbol{l}_{1}$ and $\boldsymbol{a}_{2}=m \boldsymbol{l}_{2}$, where $n,m \in \mathbb{N}$. Additionally, the atoms to be removed are highlighted in red, situated within a red circle of radius $R$. For notation purposes, we will designate this unit cell as UCHG$(n,m,R)$. c) Two-dimensional holey graphene.
			d) and e) Holey graphene nanoribbons with zigzag and armchair edges, respectively, denoted as ZHGN$(n,m,R)$ and AHGN$(n,m,R)$. Except in the case of b), sites  $A$ are denoted in blue, while sites  $B$ are represented in red	\label{fig:1-1}}
	\end{figure}

	Our model is based on periodically distributed holes within a graphene lattice, maintaining translational symmetry. This enables us to simplify the treatment by leveraging Bloch's theorem. As seen in Fig. \ref{fig:1-1} a),  we use a 4-atom superlattice unit cell containing the usual graphene non-equivalent sites $A$ and $B$, where $A$ and $B$ refer to carbon atoms on each of the graphene's bipartite lattices, distinguished in the figure by colors blue and red, respectively.
		
	Additionally, we use two orthogonal lattice vectors: $\boldsymbol{l}_{1}=( \sqrt{3} a_0,0)$ and $\boldsymbol{l}_{2}=(0,3a_0)$, where $a_0=1.42$ \r{A} is the carbon-carbon distance. As visually depicted in Fig. \ref{fig:1-1} a) and b), this specific configuration ensures the alignment of the horizontal axis with the zigzag orientation and the perpendicular axis with the armchair orientation.
	
	After obtaining the graphene lattice, we selectively remove atoms located within circles of radius $R$ (\AA) (refer to Fig. \ref{fig:1-1}c). We define the unit cell by employing lattice vectors $\boldsymbol{a}_{1}=n \boldsymbol{l}_{1}$ and $\boldsymbol{a}_{2}=m \boldsymbol{l}_{2}$ (as shown in Fig. \ref{fig:1-1}b), where $n \in \mathbb{N}$ to ensure periodicity in the $x$ direction, and similarly, $m \in \mathbb{N}$ is employed to establish periodicity in the $y$ direction. It is worth noting that, without loss of generality, the circle delineating the atoms for removal is centered at $(\boldsymbol{a}_{1}+\boldsymbol{a}_{2})/2$. As depicted in Fig. \ref{fig:1-1}, we obtain a lattice featuring holes; and cutting exclusively in the horizontal or vertical direction results in nanoribbons with zigzag or armchair boundaries, respectively (refer to Fig. \ref{fig:1-1}d and e). We will denote the unit cell of graphene with a hole of radius $R$ and vectors $\boldsymbol{a}_{1}=n \boldsymbol{l}_{1}, \boldsymbol{a}_{2}=m \boldsymbol{l}_{2}$ as UCHG$(n,m,R)$, and correspondingly, the nanoribbons with zigzag and armchair directions as ZHGN$(n,m,R)$ and AHGN$(n,m,R)$, respectively.

	Under such considerations, our investigation is based on a tight-binding model with only first-neighbors hopping transfer integral, which yields a Hamiltonian given by
	\begin{equation}\label{eq:tigh-binding-hamiltonian}
		\mathcal{H}= \sum_{\langle ij \rangle} t_{0} \hat{c}^{\dag}_{\boldsymbol{r}_{i}}\hat{c}_{\boldsymbol{r}_{j}} + h.c
	\end{equation}
	where $\langle ij \rangle$ represents the sum over the neighbors with positions $\boldsymbol{r}_{i}$ and $\boldsymbol{r}_{j}$ that satisfy $|\boldsymbol{r}_{j}-\boldsymbol{r}_{i}|=a_0$; $\hat{c}^{\dag}_{\boldsymbol{r}_{i}} (\hat{c}_{\boldsymbol{r}_{i}})$ is the creation (annhilation) operator and $t_0=-2.8$ eV is the hopping integral between the $i$-th and $j$-th sites. Additionally, we numerically construct the Hamiltonian operator in  reciprocal space $\boldsymbol{k}$, which depends on the number of atoms in the unit cell and is denoted as $\hat{\mathcal{H}}_{T}(\boldsymbol{k})$. The eigenvalues and eigenfunctions were thus numerically found by using python dedicated libraries \cite{pybinding,  pyqula, Tbplas2023}. In the following section we will discuss the resulting electronic and optical properties.

	\section{Electronic properties of bidimensional Holey Graphene \label{sec: electronic properties}}
	
	\begin{figure}[t]
		\fl
		a)\includegraphics[height=0.28\textheight]{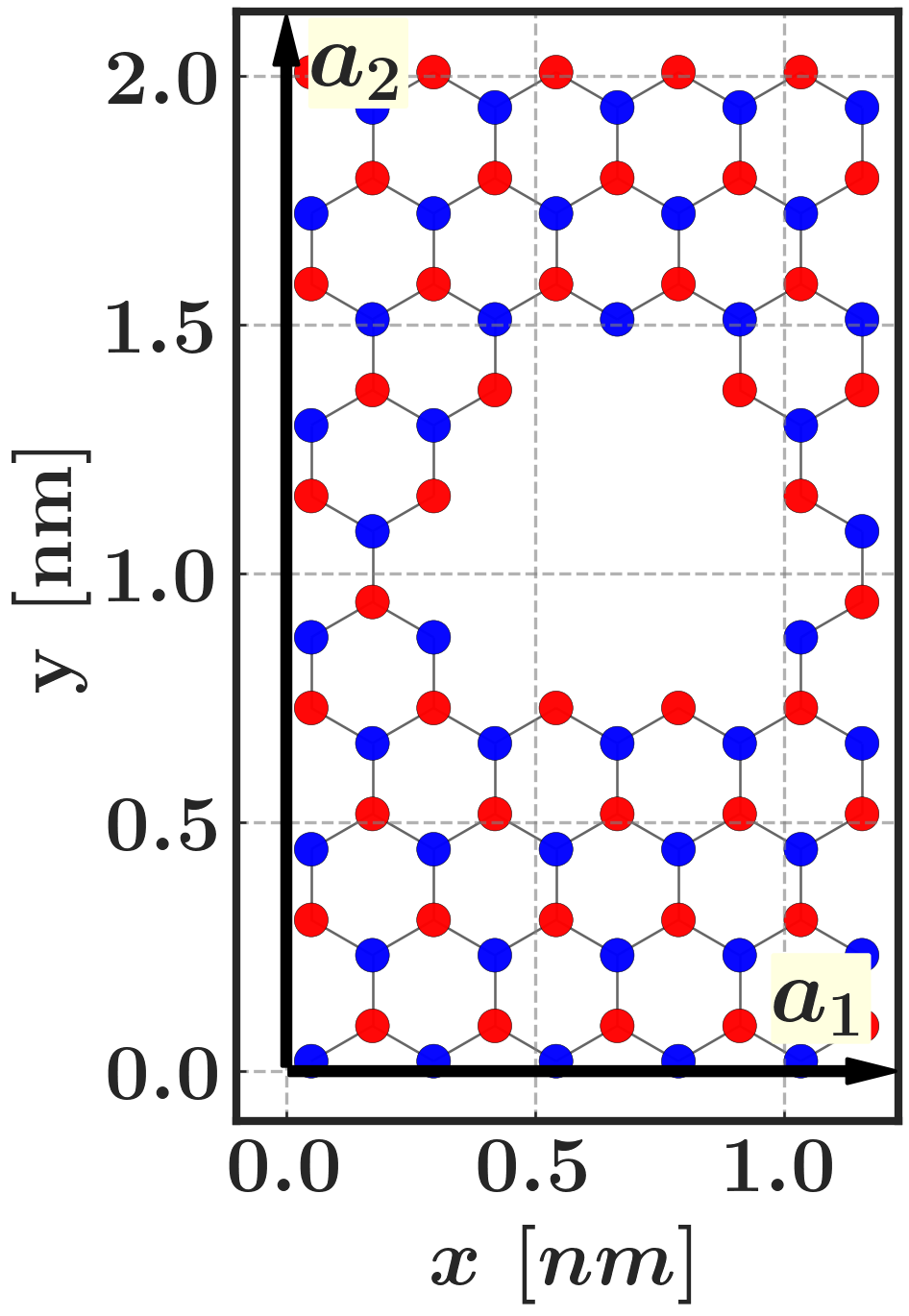}
		b) \includegraphics[height=0.28\textheight]{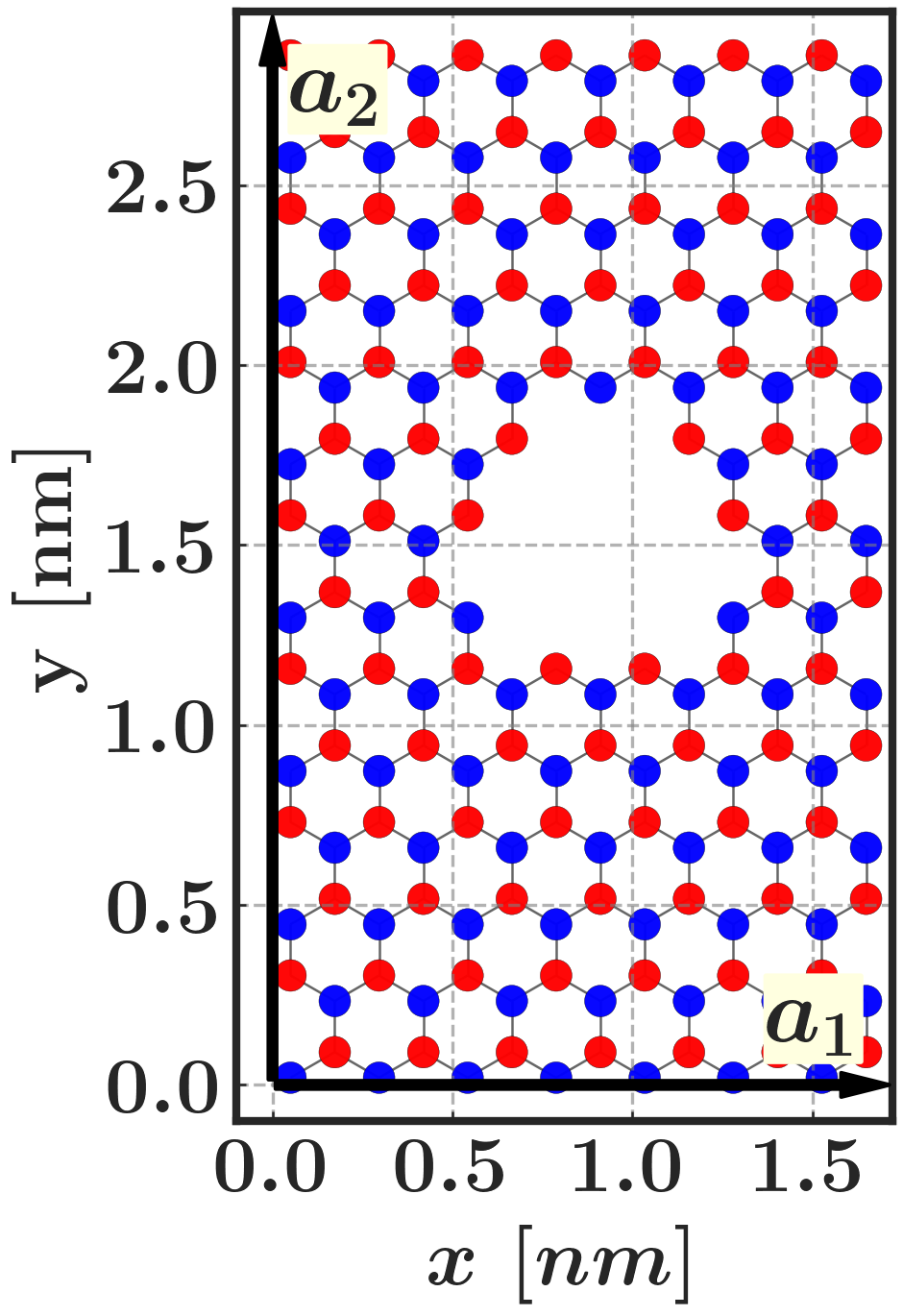}
		c) \includegraphics[height=0.28\textheight]{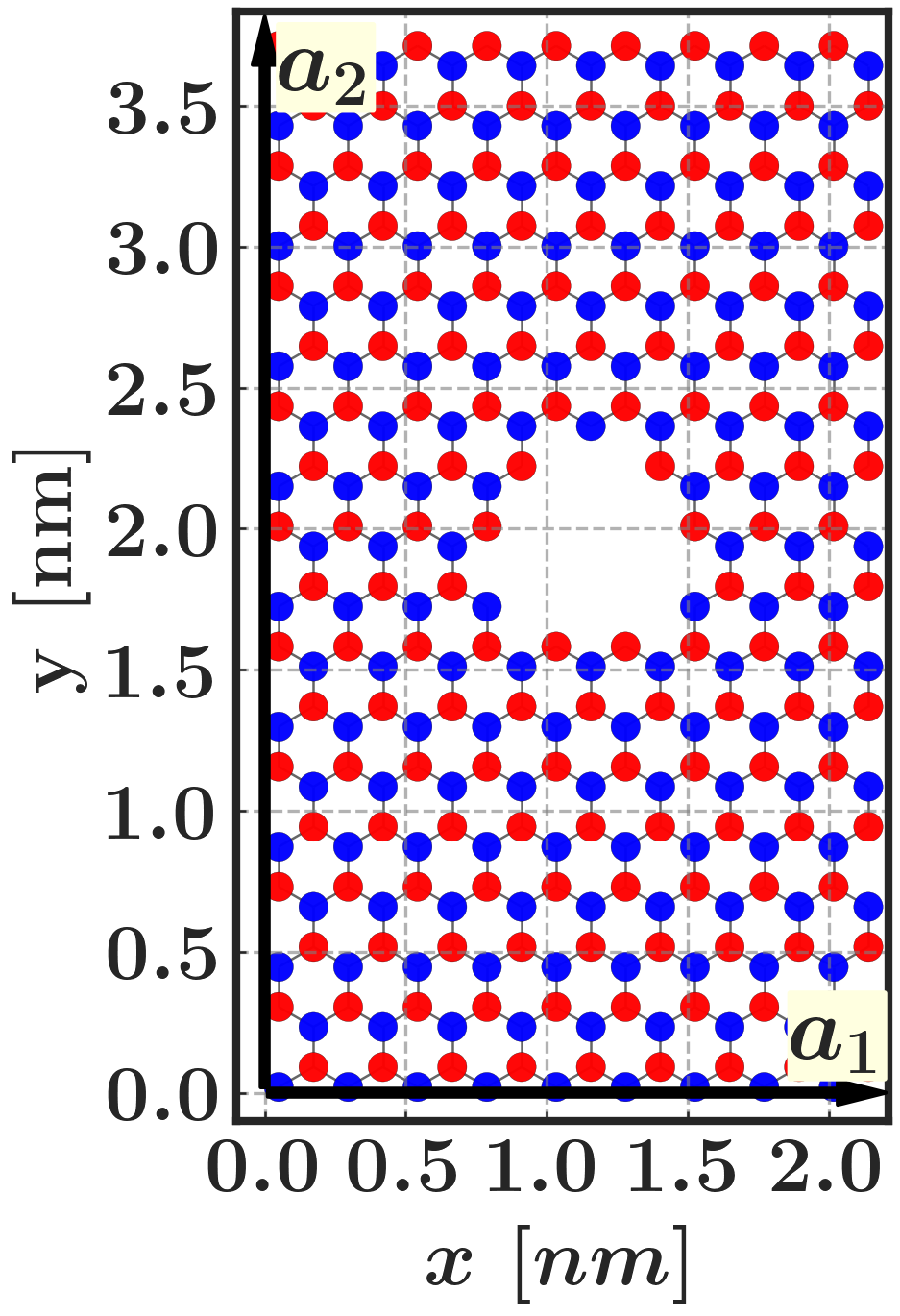}\\
		\fl
		d) \includegraphics[height=0.28\textheight]{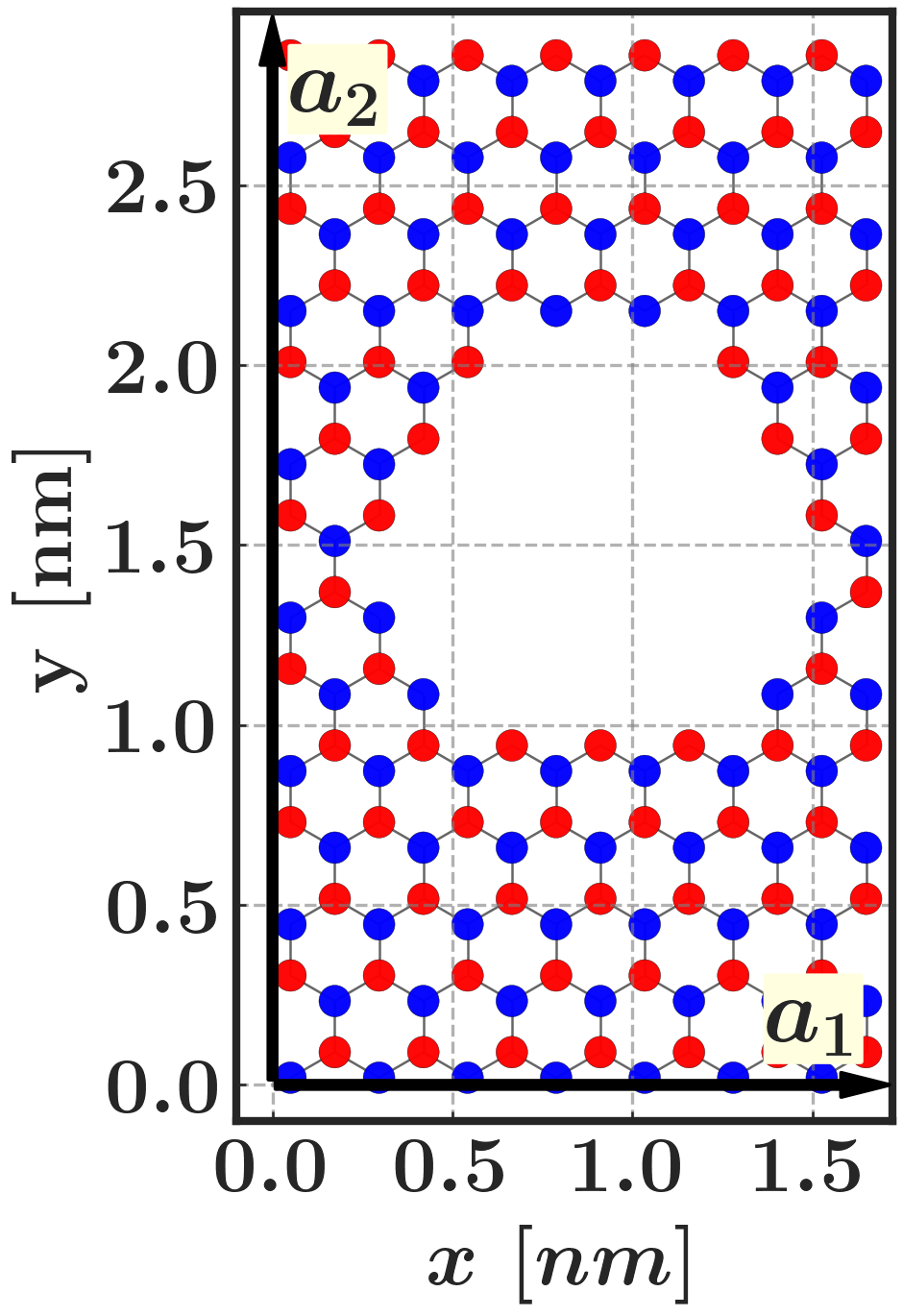}
		e) \includegraphics[height=0.28\textheight]{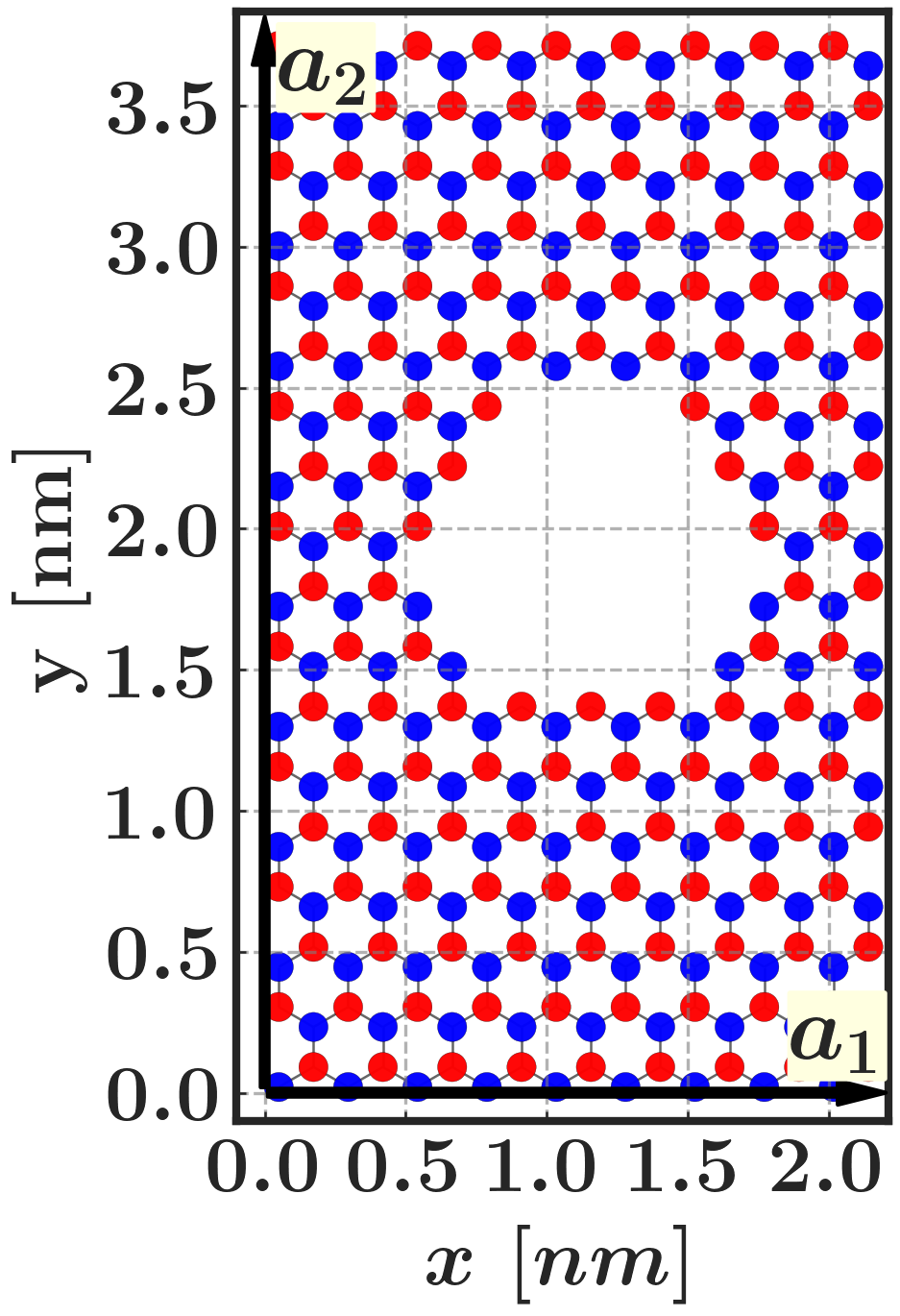}
		f) \includegraphics[height=0.28\textheight]{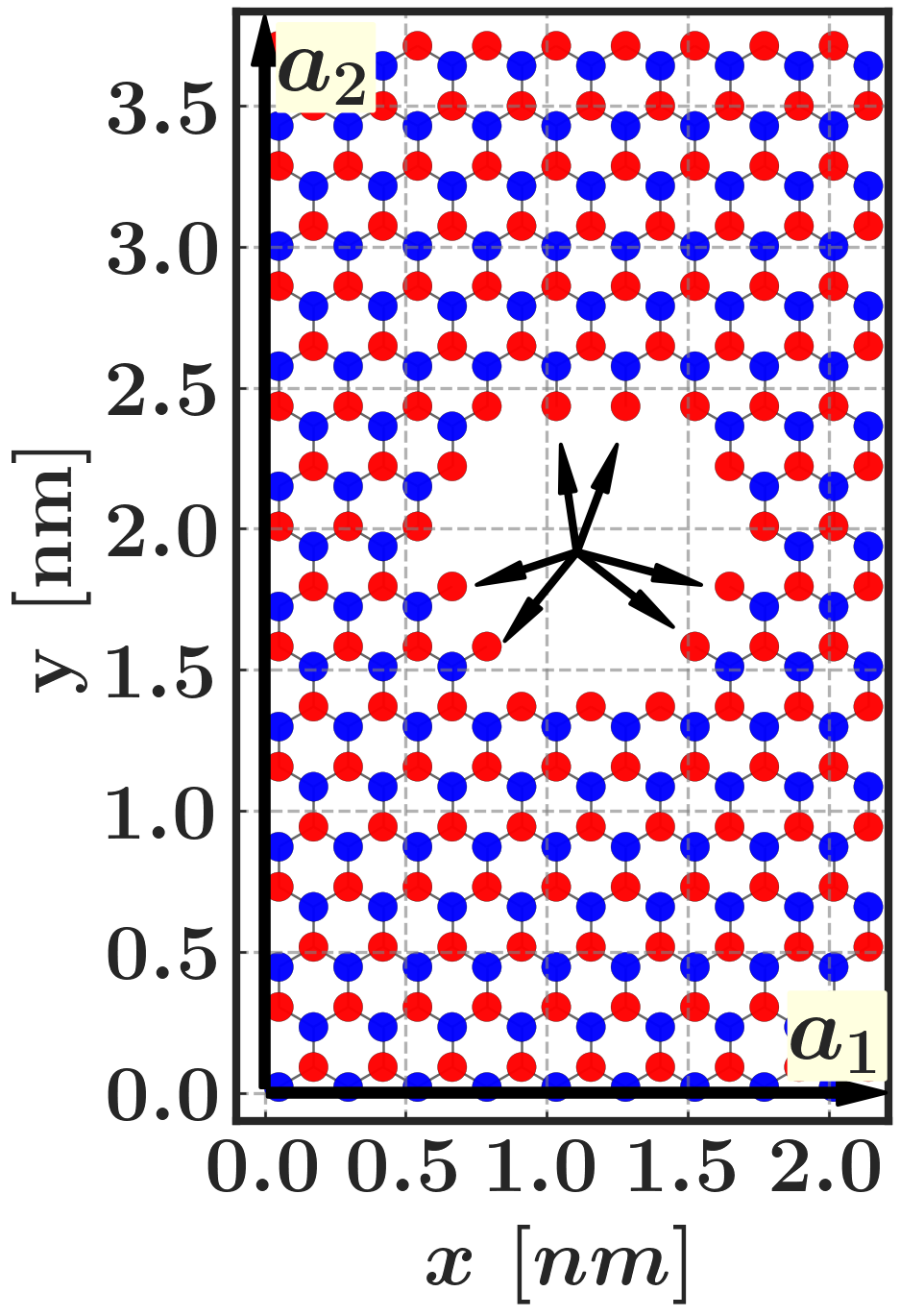}
		\caption{Six unit cells are displayed in which we vary the radius and the size of the corresponding lattice vectors for a) UCHG$(5,5,3)$, b) UCHG$(7,7,3)$, c) UCHG$(9,9,3)$, d) UCHG$(7,7,5.2)$, e) UCHG$(9,9,5.2)$, and f) UCHG$(9,9,5.0)$. This last case can be considered as the scenario for $UCHG(9,9,5.2)$ with the addition of six dangling atoms with $C_3$ symmetry are observed. Recently, it has been shown that incorporating dangling bonds with oxygen atoms modulates the active sites of holey graphene, altering its chemical reactivity \cite{Koh2023}. The carbon atom colors indicate sites on each of the bipartite sublattices $A$ and $B$. \label{fig:2-1}}
	\end{figure}

\begin{figure}[t]
		\fl
		a) \includegraphics[scale=0.4]{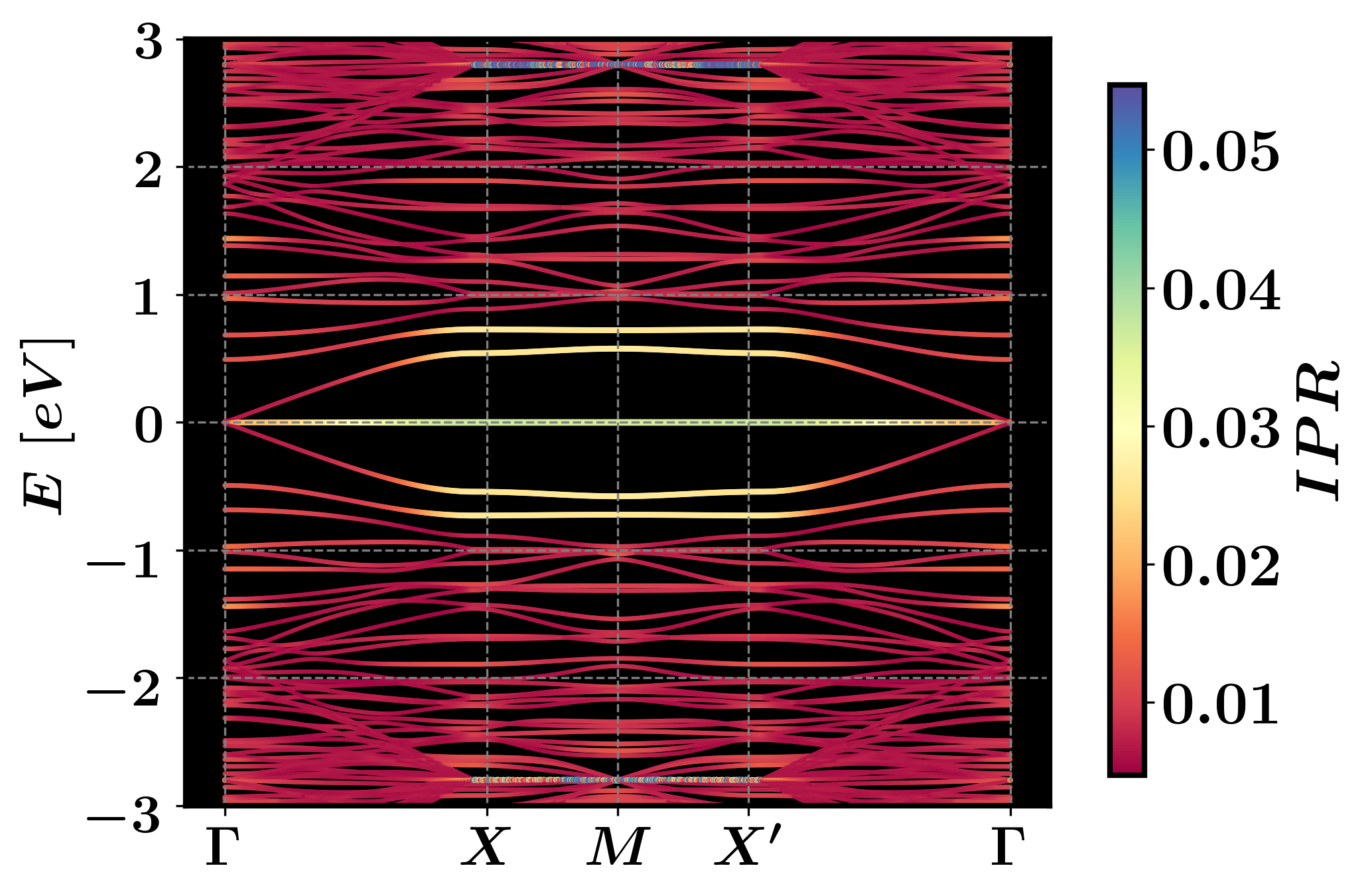}
		b)\includegraphics[scale=0.4]{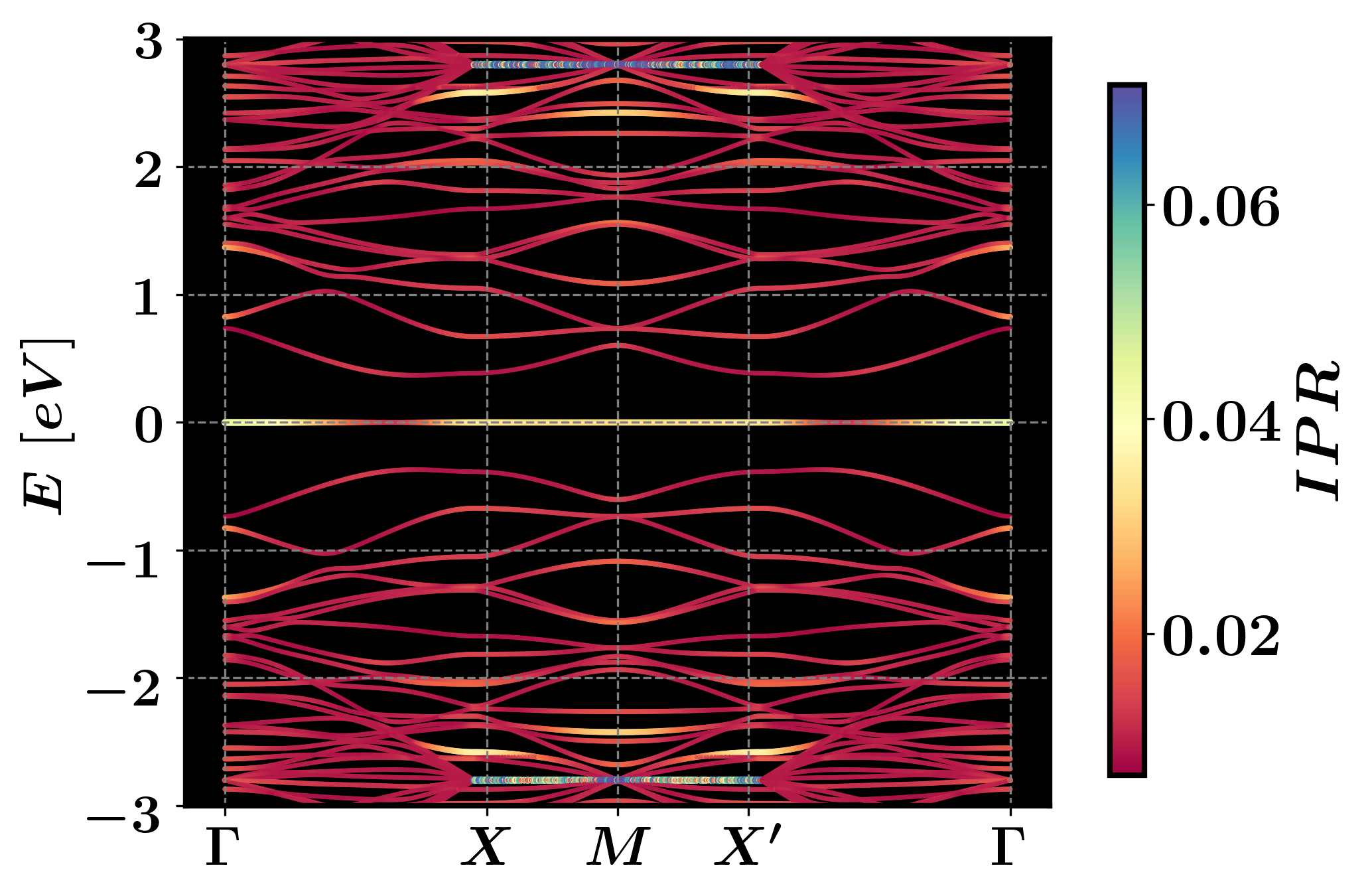}\\
		\fl
		c)\includegraphics[scale=0.4]{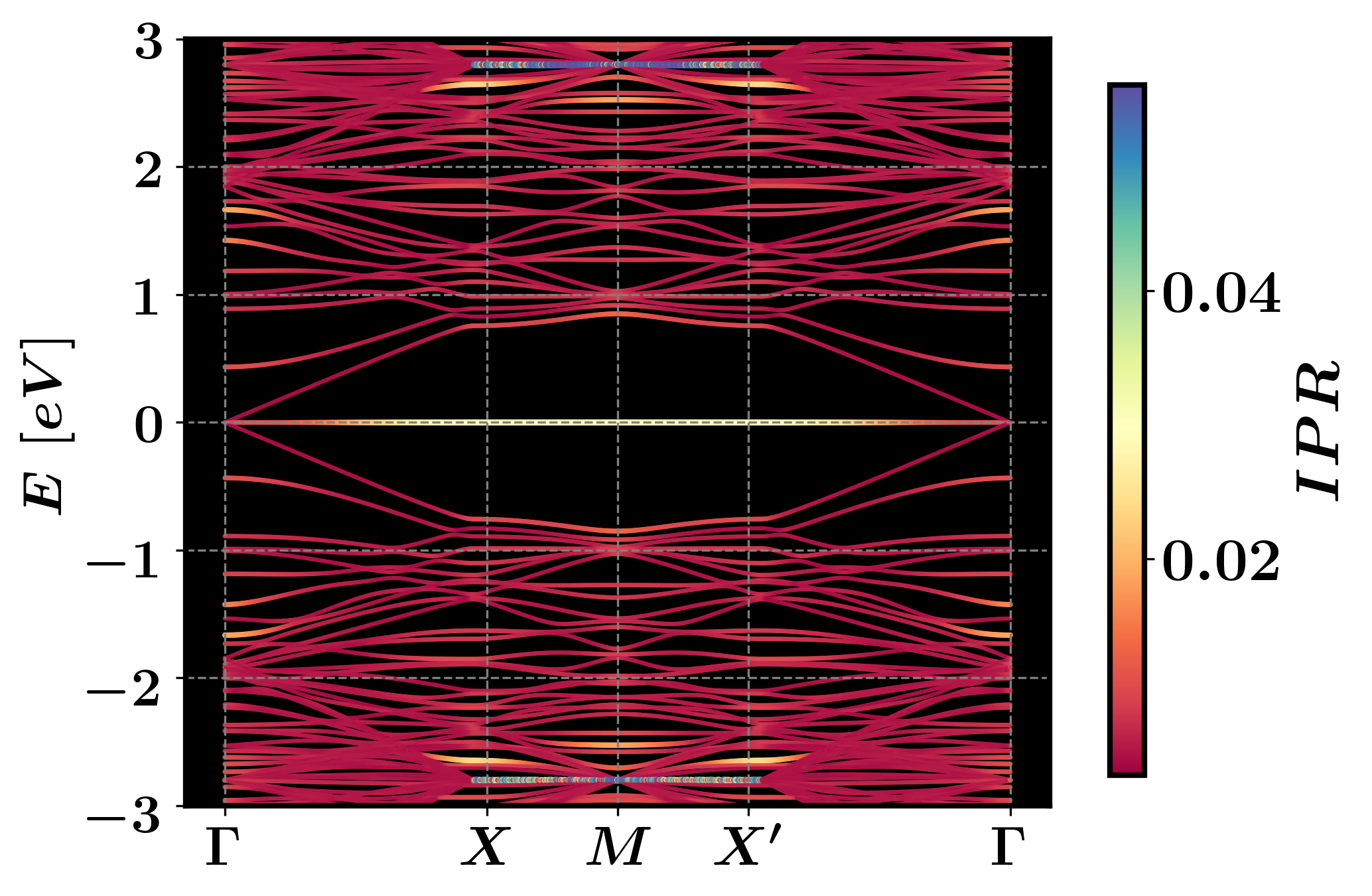}
		d)\includegraphics[scale=0.4]{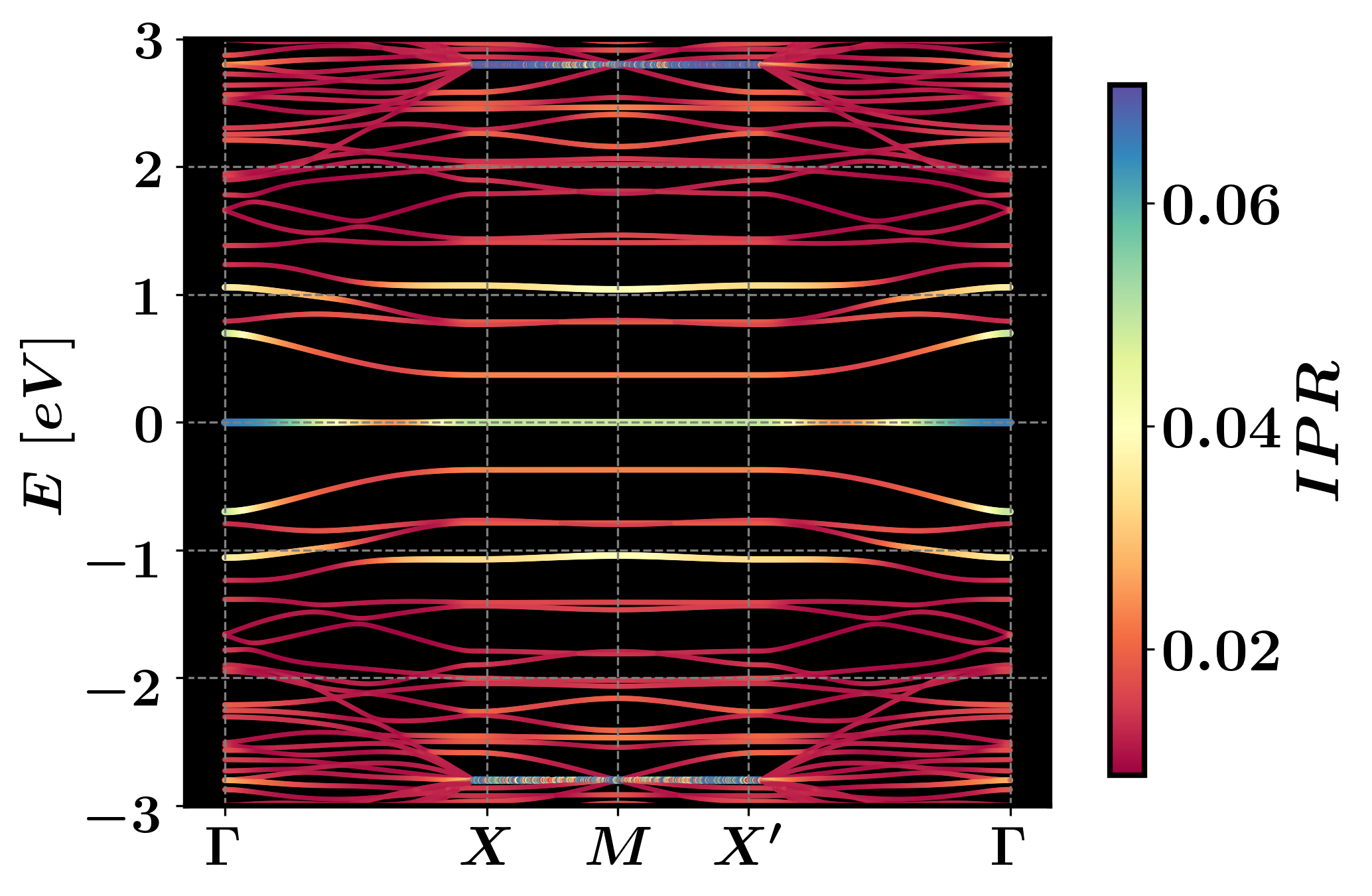}\\
		\fl
		e)\includegraphics[scale=0.4]{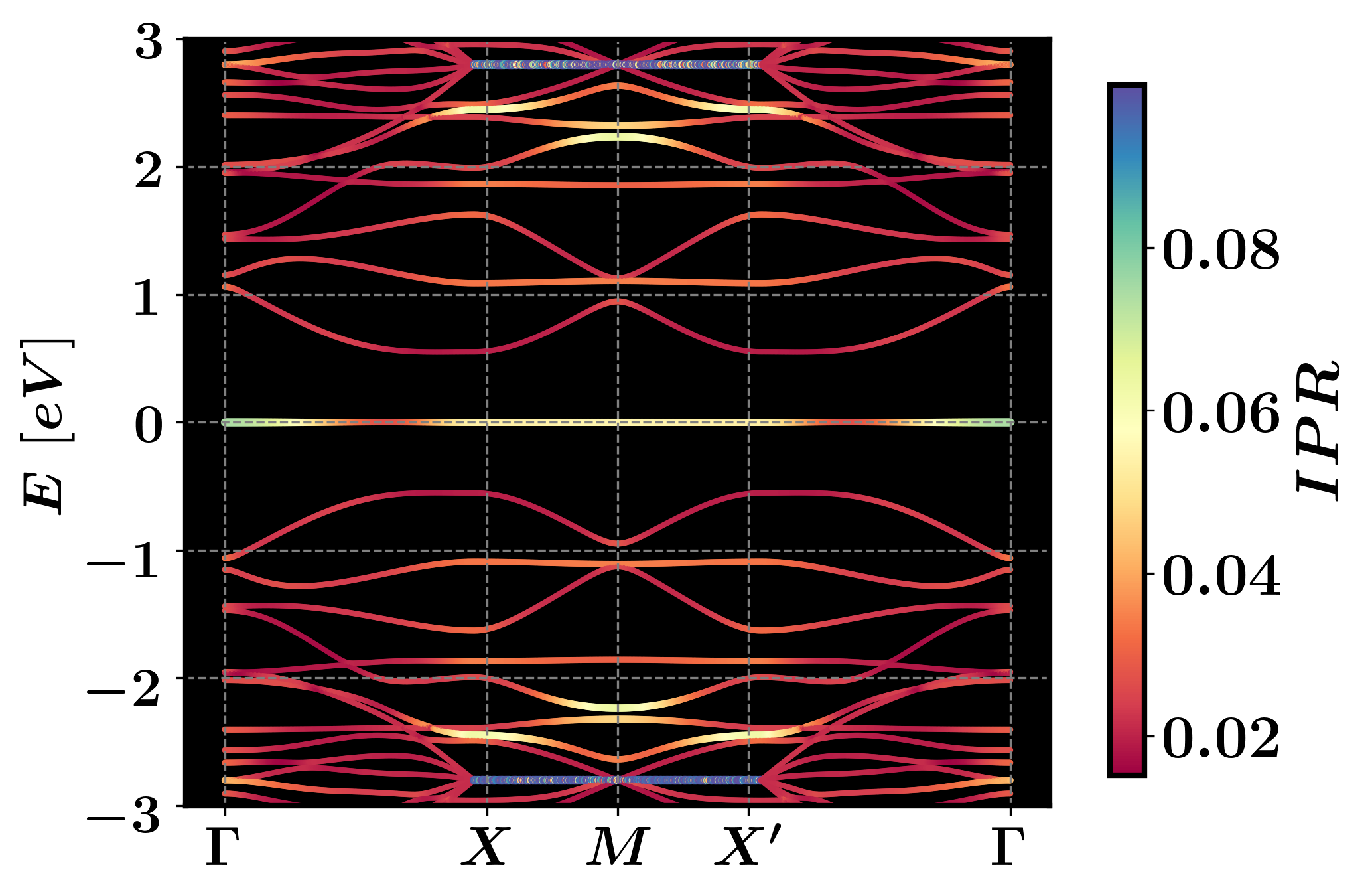}
		f)\includegraphics[scale=0.4]{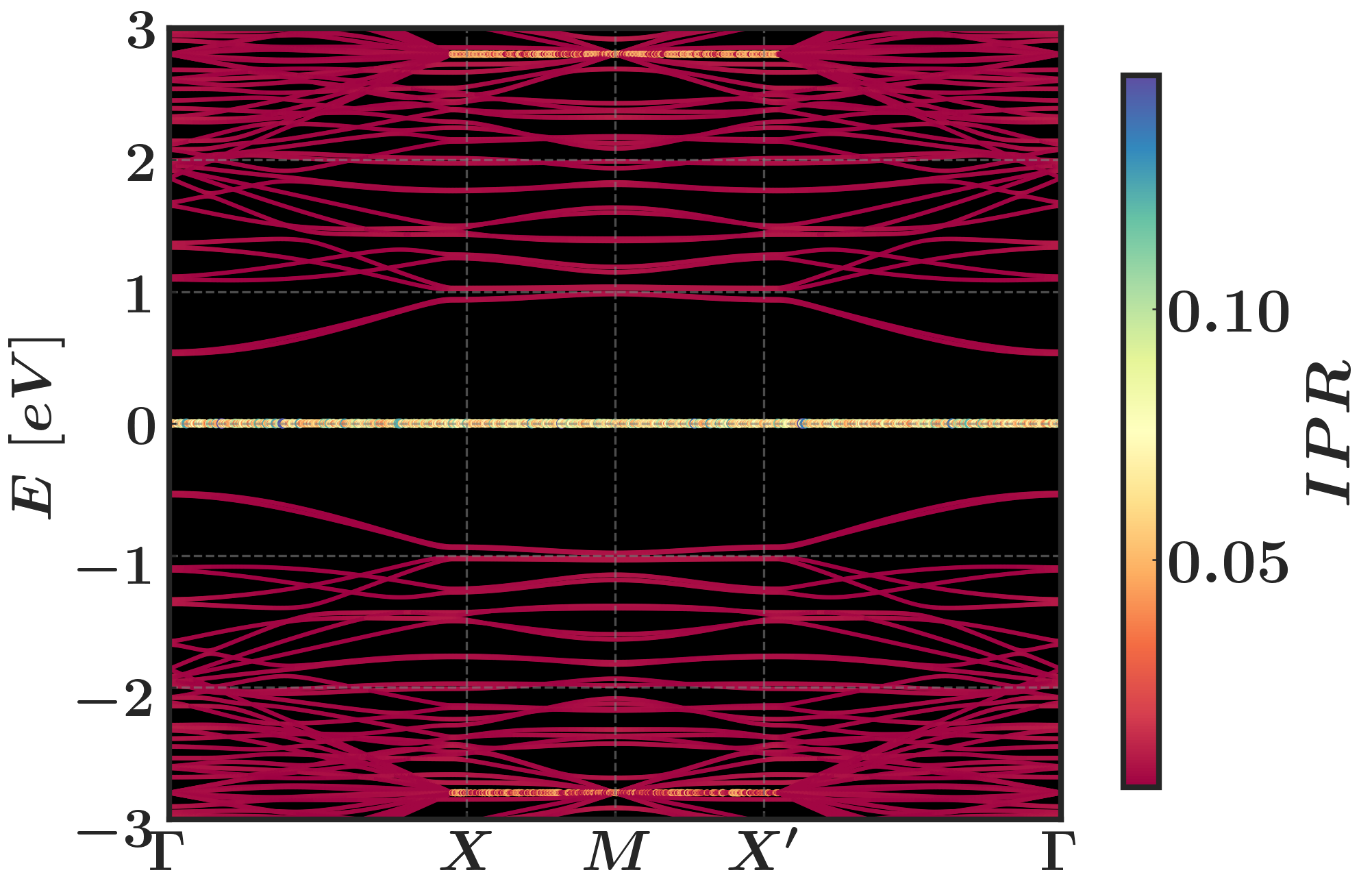}
		\caption{The band structure for systems with unit cells of a) UCHG$(5,5,3)$, b) UCHG$(7,7,3)$, c) UCHG$(9,9,3)$, d) UCHG$(7,7,5.2)$, e) UCHG$(9,9,5.2)$, and f) UCHG$(9,9,5.0)$ are shown. It is observed that the states in the flat bands exhibit characteristics of Compact Localized States (CLS). Furthermore, for cases a) to d), it is noted that as $n$ increases while maintaining $R$, the localization decreases; however, this does not hold true for case f). This behavior can be understood in connection with the inbalance density $\rho(n,R)$, as discussed in the text. On the other hand, a consequence of maintaining dangling bonds distributed with $C_3$ symmetry is the creation of a gap. This gap remains strongly protected even when increasing the size of the cell.
			\label{fig:3}}
	\end{figure}
 
In this section, we will discuss the electronic and optical properties of systems with unit cells of type UCHG$(n,n,R)$, focusing primarily on those depicted in Figure \ref{fig:2-1}, namely, UCH$(5,5,3)$, UCHG$(7,7,3)$, UCHG$(9,9,3)$, UCHG$(7,7,5.2)$, UCHG$(9,9,5.2)$, and UCHG$(9,9,5)$.
These examples were chosen to include the effects of different radius and edge terminations. Observe that among our examples, in  Fig. \ref{fig:2-1} f) we include the case of a hole with dangling bonds.

In Fig. \ref{fig:3} we illustrate the bands obtained for each of the systems shown in Fig. \ref{fig:2-1}. 
As we can see, gaps are observed in some cases while Dirac cones are seen in others. The opening of these gaps will be discussed later on in this section. Meanwhile, we observe that  at  energy $E=0$ flat bands are obtained in all cases, corresponding to the Fermi energy ($E_F = 0$ eV) at half-filling. Recent research suggests that the formation of flat bands is intricately linked with Compact Localized States (CLS) \cite{Espinosa-Champo_2024}, also known as confined states \cite{Naumis1994,Naumis2002}, signifying the existence of a non-trivial localization behavior. 
	
In order to ascertain whether there exists differences in the localization properties of states within the flat bands and those outside of them, we employ the Inverse Participation Ratio (IPR) \cite{Bell1970,Edwards1972,Shukla2018, Naumis2007, Wegner1980}
	
	\begin{equation} \label{eq: eq-IPR}
		IPR(E(\boldsymbol{k},s))= \int_{UC} |\psi_{\boldsymbol{k},s}(\boldsymbol{r})|^{4} d^{2} r
	\end{equation}

where $\psi_{\boldsymbol{k},s}(\boldsymbol{r})$ represents the wave function in the $s$-th bands and characterized by momentum $k$ and energy $E(\boldsymbol{k},s)$, and due to the periodic condition of the lattice, we perform the integral over the unit cell (UC). The IPR of an extended state goes as $1/N$ where $N$ is the number of extended states, while for localized states does not depend on $N$. For exponentially localized states, it can be proven that  $IPR(E) \sim \lambda^{-2}$ where $\lambda$ is the localization length,.

\begin{figure}[t]
		\fl
		a)	\includegraphics[width=0.35\textwidth]{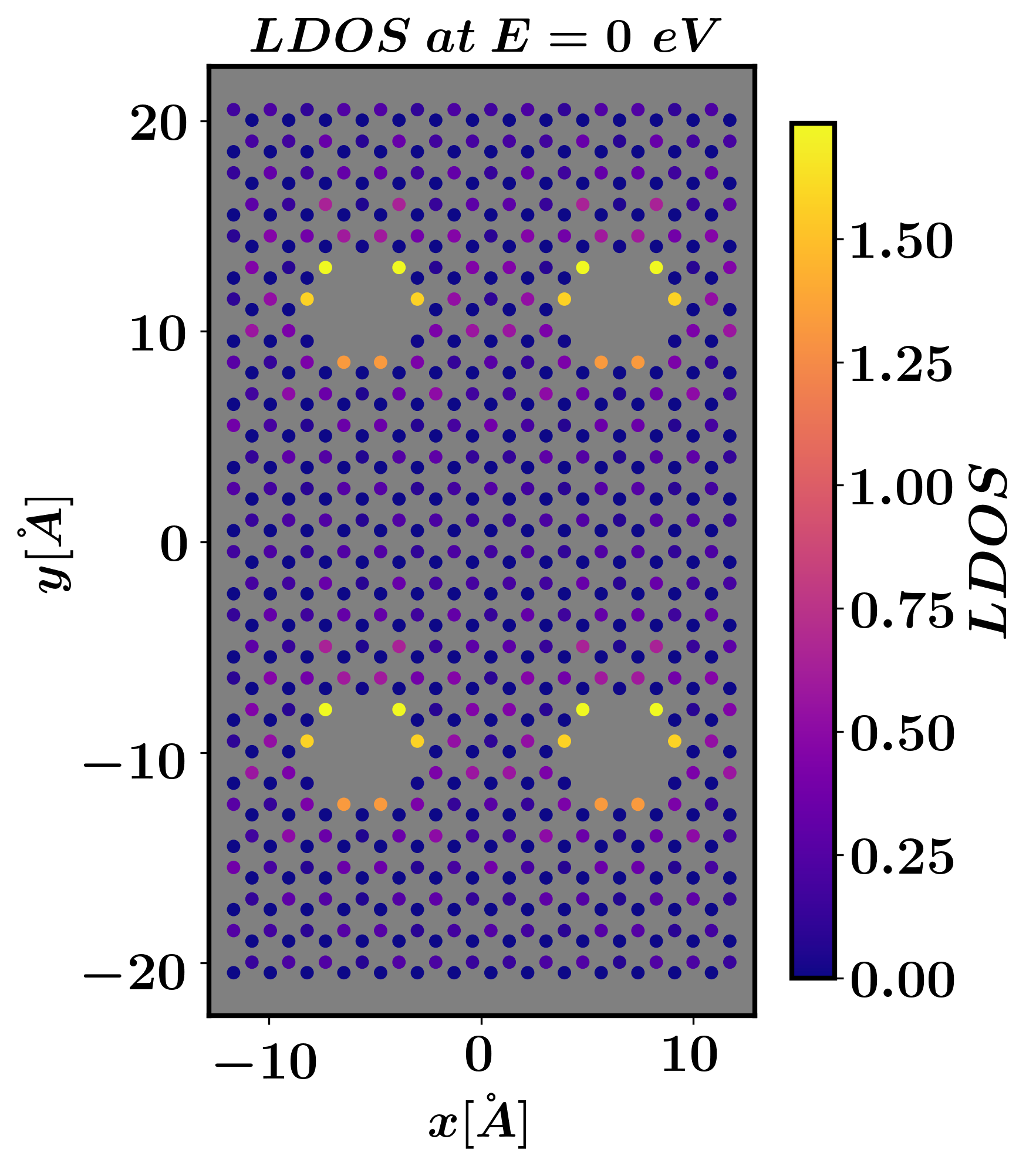}
		b)\includegraphics[width=0.35\textwidth]{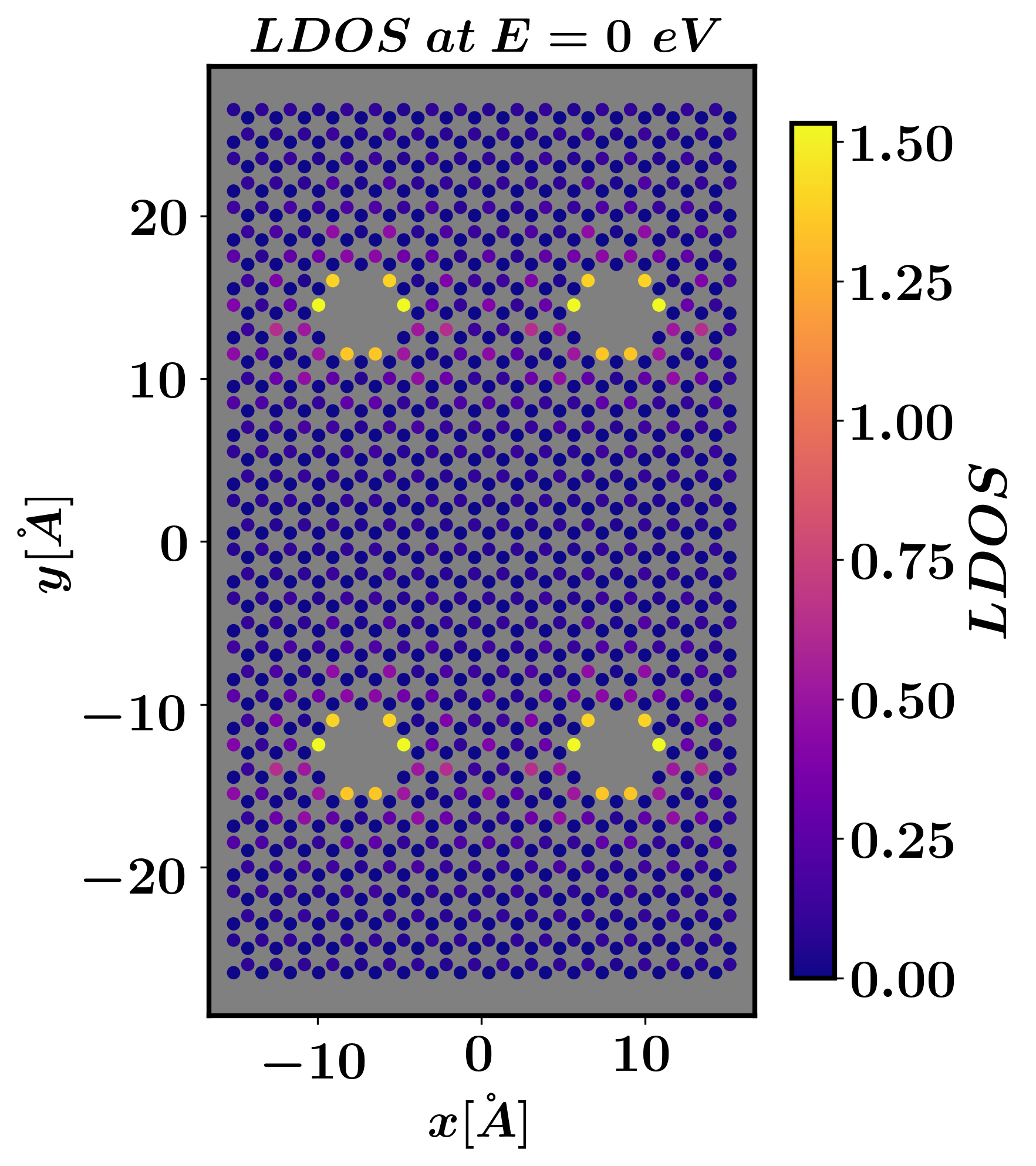}
		c) \includegraphics[width=0.4\textwidth]{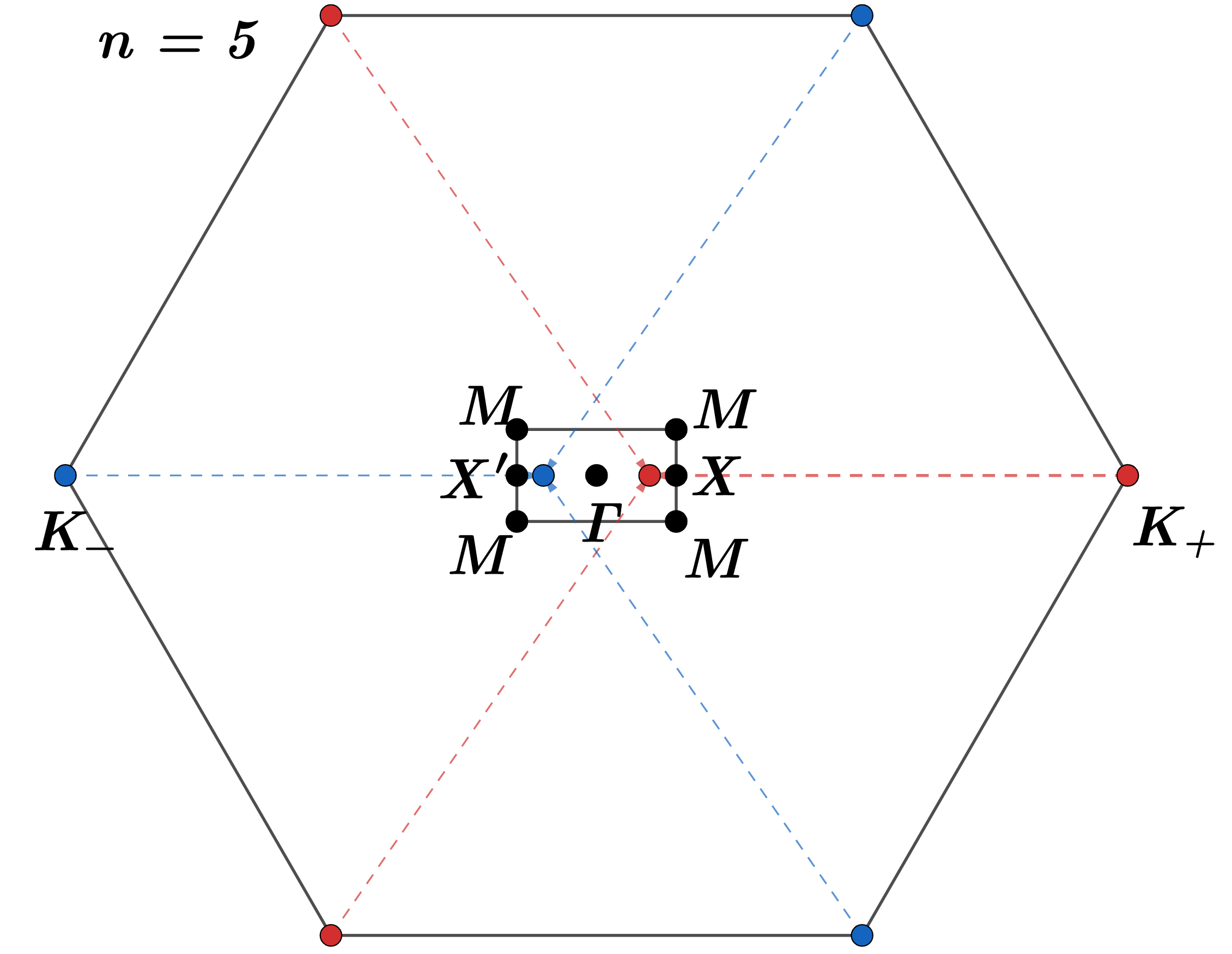}\\
		\fl
		d)	\includegraphics[scale=0.35]{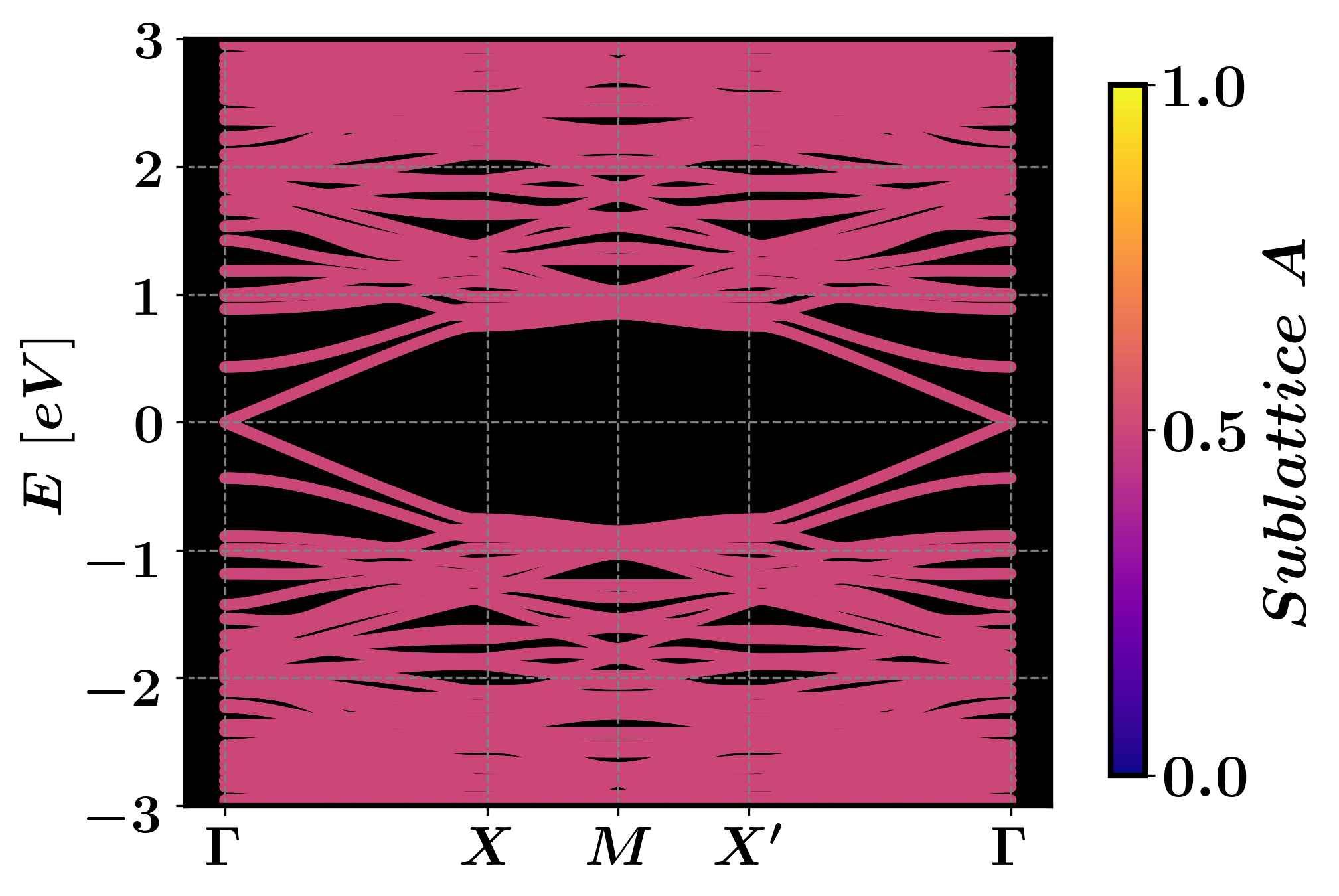}
		e)	\includegraphics[scale=0.35]{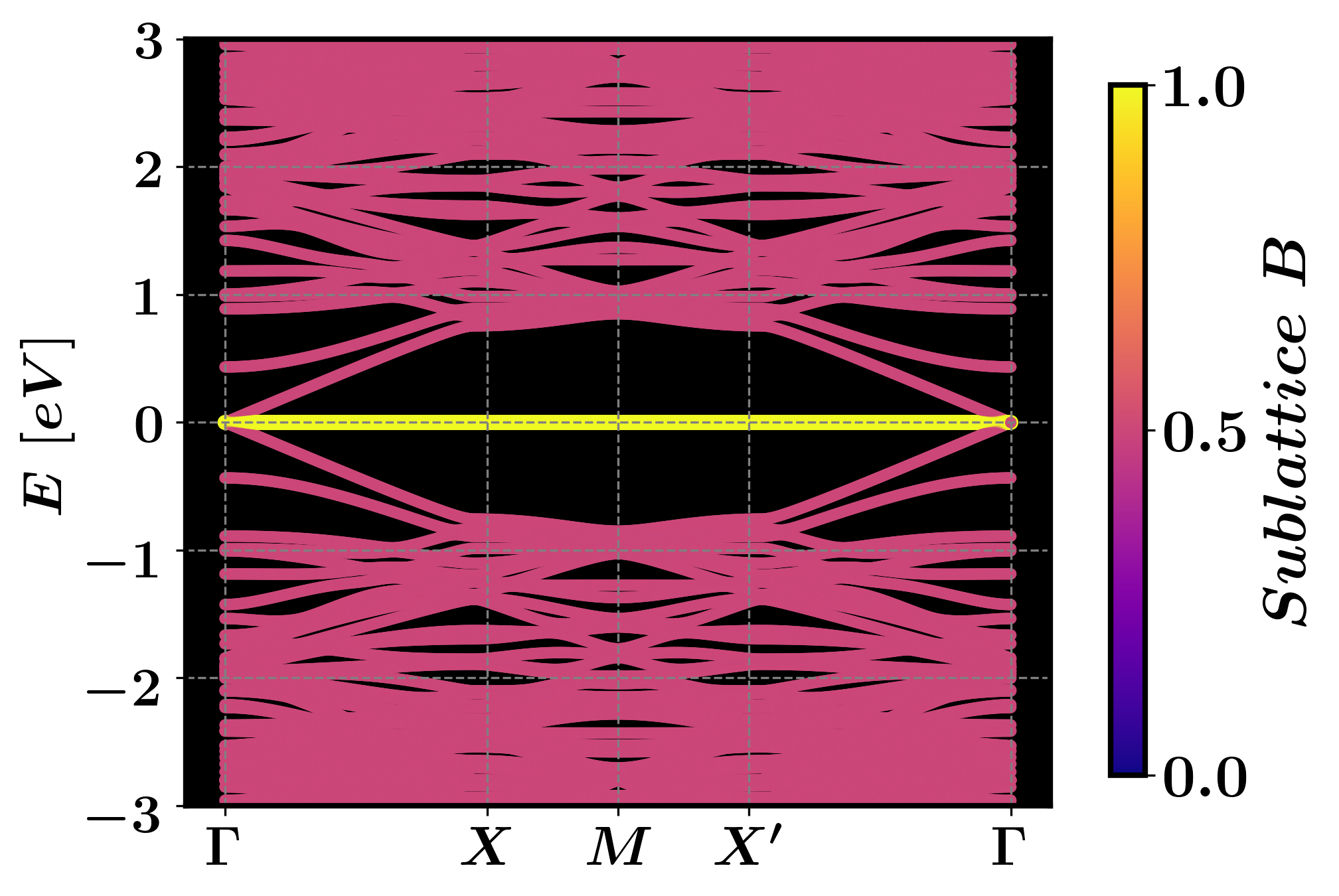}\\
		\fl
		f)	\includegraphics[scale=0.4]{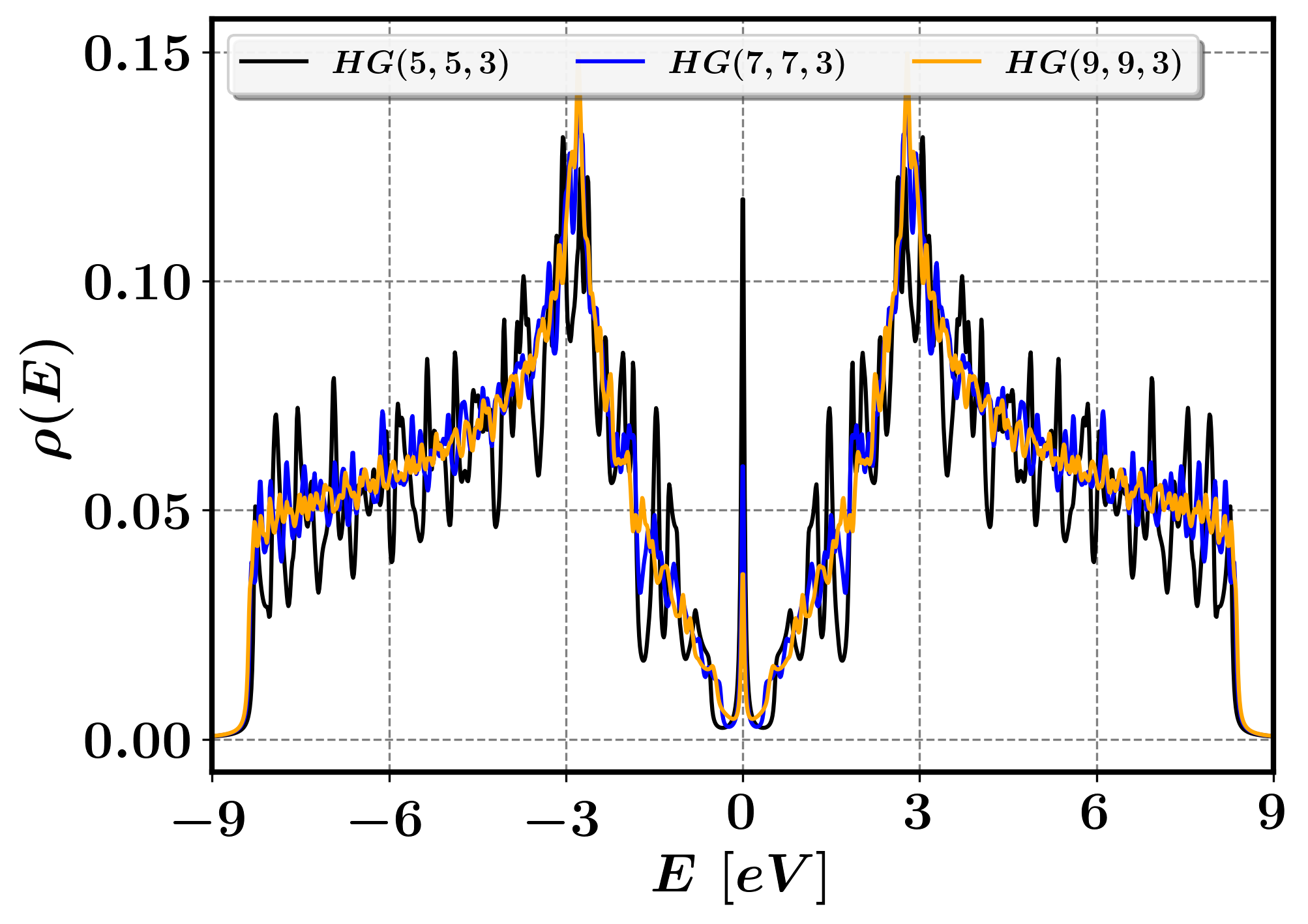}
		g)	\includegraphics[scale=0.4]{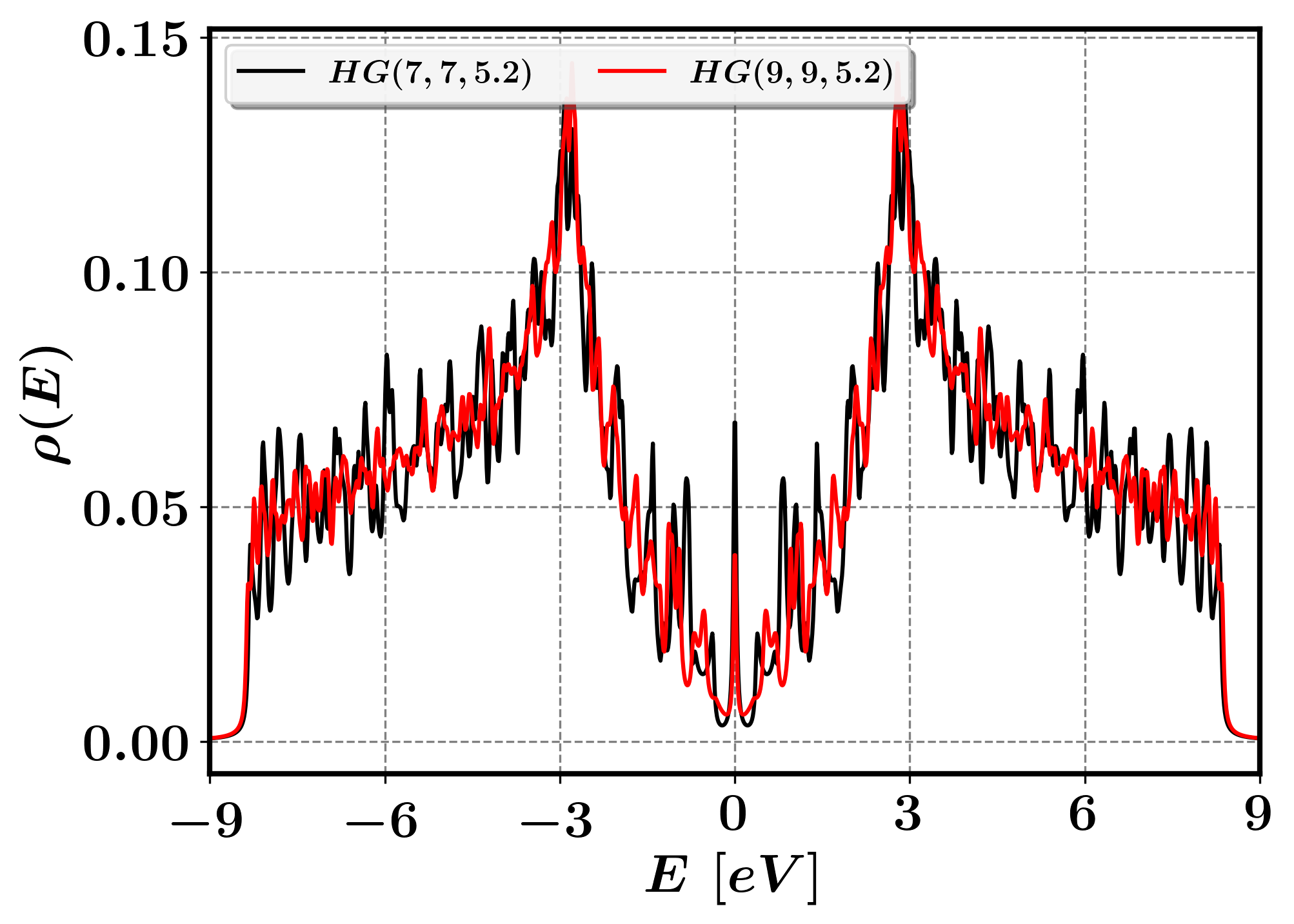}
\caption{ a)-b)The spatial Local Density of States (LDOS) for the UCHG$(7,7,3)$ and UCHG$(9,9,3)$ cells at $E=0$ is displayed, with the states corresponding to the flat band being associated with edge CLS on the zigzag boundaries of the holes. c) The hexagonal Brillouin zone (BZ) of graphene is folded into the rectangular BZ of the HG. d)-e) The contribution of sublattice states in the band structure of UCHG$(9,9,3)$ is shown, with the flat band resulting from the imbalance number through the path exchange symmetry \cite{Jun2023, Espinosa-Champo_2024}. f)-g)The density of states, $\rho(E)$, is presented for the unit cells f)UCHG($5,5,3$) (black line), UCHG($7,7,3$) (blue line), UCHG($9,9,3$) (orange line), g) UCHG($7,7,5.2$) (black line), and UCHG($9,9,5.2$) (red line). It is observed that $\rho(E=0)$ decreases proportionally to the size of the cell $n$, which is a consequence of the imbalance density $\varrho(n,R)$ as established in Table \ref{tab:table1} and discussed previously.  \label{fig:4}}
	\end{figure}

In Figure \ref{fig:3}, the $IPR(E(\boldsymbol{k},s))$ values for each $E_{\boldsymbol{k,s}}$ are projected onto the band structure using a color code. As anticipated, the flat band states exhibit a significantly high degree of localization in contrast to the majority of other states. Flat band states preferentially spatially localize in the zig-zag boundary regions that form around the hole, as can be observed in Figure \ref{fig:4} a) and b). However, it is worth noting that these localized states are not the sole examples; in addition, we find states for which the bands exhibit quasi-flat behavior within specific regions of the Brillouin zone, notably those with energies at $E=|t|$ and momenta between the $X$ and $M$ points. Such behavior is due to the presence of Van Hove singularities in which a dimerization effect is seen \cite{Leonardo2021, Gerardo2014, Elias2023}.

To further confirm the nature of localization, in Fig. \ref{fig:3} a)-b) we present the local density of states (LDOS) for a flat band state as a function of the sites for two different examples. Therein it can be seen that localization mainly occurs at the edges of the holes as expected. Also, from Figs. \ref{fig:3} a)-d) and \ref{fig:4} a) and b), it is observed that if $n$ increases while $R$ remains constant, or vice versa, the localization in the flat band decreases. However, for $n=9$ and $r=5$ such rule does not to hold since in addition to a increased localization, a gap opens. To gain a better understanding on the formation of flat bands, the gap size, $\Delta$ and the density of states at $E=0$, we introduce two parameters. The first one is the {\it imbalance number}, denoted as $\delta N(n,R)$, given by
	\begin{equation} \label{eq:eq-imbalance number}
		\delta N(n,R)=|N_{A}-N_{B}|
	\end{equation}
which indicates the difference between number of sites $A$ and $B$ in the unit cell UCHG$(n,n,R)$. The second is the {\it imbalance density}, $\varrho(n,R)$, given as
	
	\begin{equation} \label{eq:eq-imbalance-density}
		\varrho(n,R)= \delta N(n,R)/N,
	\end{equation}
	where $N$ is the total number of atoms in each unit cell.
	
As established in Ref.  \cite{BARRIOSVARGAS201323}, sublattice site imbalance in bipartite lattices can induce flat bands of confined states, and in certain cases, the formation of energy gaps and pseudogaps. 
Bipartite sublattice site imbalance will occur here only at hole edges. To see if such effect is in play , in Table \ref{tab:table1} we present the values for the site imbalance number, the site imbalance density, the energy gap, the density of states (DOS) at $E=0$ and the observed number of flat bands.	
	
From Table \ref{tab:table1}, we conclude that the number of flat bands formed at energy $E=0$ eV coincides with the imbalance number for all the studied cases.  In recent studies, it has been demonstrated that the formation of degenerate flat bands is also a consequence of a broken path-exchange symmetry \cite{Jun2023, Espinosa-Champo_2024}. Upon breaking, this symmetry introduces a complex phase \cite{Jun2023, Espinosa-Champo_2024}, leading to the lifting of the degeneracy while simultaneously preserving the flatness of the band. 
In the periodic holey system presented here, it can be demonstrated that the number of imbalances is related to the path exchange symmetry, and consequently the number of flat bands is equal to the imbalance number.  Specifically, when $\delta N(n,R) \neq 0$ such symmetry is broken and an isolated flat band emerges. Therefore, such symmetry distinguishes the sublattice with the highest number of sites. This is illustrated in Fig. \ref{fig:4}, panels d) and e), for the case of HG$(9,9,5.2)$, where the nonzero energy bands result from equal contributions from sublattices $A$ and $B$.

Additionally, it is worth noting that the density of states (DOS or $\rho(E)$), at $E=0$ is directly proportional to the imbalance density (see Table \ref{tab:table1}). In cases where there are no dangling bonds, we observe that $\rho(E=0) \sim 10.5 \varrho(n,R)$, while in cases with dangling bonds, we find that $\rho(E=0) \sim \varrho(n,R)$ (see Fig. \ref{fig:4} f) and g)).   This can be understood by considering the  DOS for each unit cell, in this case, $\rho(E=0) \sim n_{st}/N$, where $n_{st}$ is the number of sites supporting the state with $E=0$. Therefore, $\rho(E)/\varrho (n,R) \sim n_{st}/ \delta N(n,R)$. In the studied cases without dangling bonds, $n_{st}$ fluctuates between $9-12$ sites (see for example Fig. \ref{fig:4} a) and b)), and $\delta N(n, R)=1$, maintaining an average ratio of $10.5$ (see Table \ref{tab:table1} and Fig. \ref{fig:4} f) and g) ). For cases with dangling bonds, the $E=0$ state is strongly localized on these dangling sites, and $n_{st}$ corresponds to this number of sites, hence $\rho(E=0) \sim \varrho(n,R)$.

	\begin{table}
		\centering
		\begin{tabular}{|c|c|c|c|c|c|c|c|}
			\hline
			$n$ & $R$[\AA]  & $\delta N(n,R)$ & $\varrho(n,R)$ & $\Delta$ [eV] &   $\rho(E=0)$  & No. Flat Bands \\
			\hline
			$5$ & $3$ & $1$ & $\sim 1.15\times 10^{-2}$  & $\sim 0.551$ & $\sim 0.119$  & $1$  \\
			\hline
			$7$ & $3$ & $1$  & $\sim 5.46 \times 10^{-3}$ &  $\sim 0.369$ &  $\sim 0.060$ & $1$   \\
			\hline
			$9$ & $3$ & $1$ & $\sim 3.21 \times 10^{-3}$ & $\sim 0.000$ & $\sim 0.0364$  & $1$  \\
			\hline 
			$7$ & $5.2$ & $1$  & $\sim 6.30 \times 10^{-3}$ & $\sim 0.371$ & $\sim 0.0684$  & $1$  \\
			\hline 
			$9$ & $5.2$ & $1$ & $\sim 3.50 \times 10^{-3}$ &$\sim 0.000$  & $\sim 0.0402$ & $1$  \\
			\hline
			$9$ & $5$  & $7$ & $\sim 2.39 \times 10^{-2}$ & $\sim 0.528$ & $\sim 0.024$ & $7$   \\
			\hline
		\end{tabular}
		\caption{Table displaying the number of imbalances, $\delta N(n,R)$, the imbalance density, $\varrho(n,R)$, gap size, $\Delta$, state density, $\rho(E)$, and the number of flat bands for graphene with a gap of size $R$ [\AA] and unit cells UCHG$(n,n,R)$.}
		\label{tab:table1}
	\end{table}

\begin{figure}[t]
\fl
a)	\includegraphics[scale=0.3]{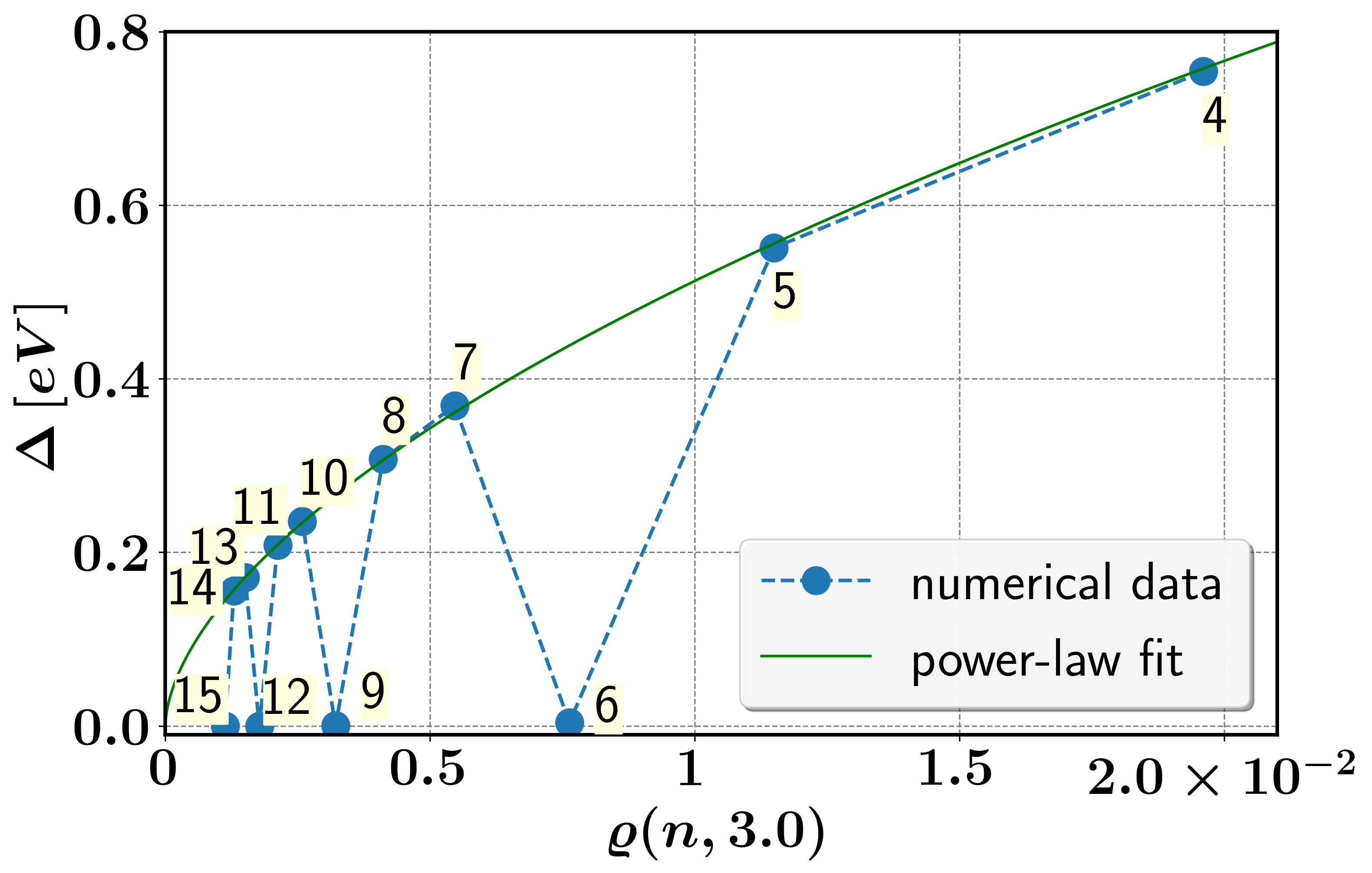}
b)\includegraphics[scale=0.3]{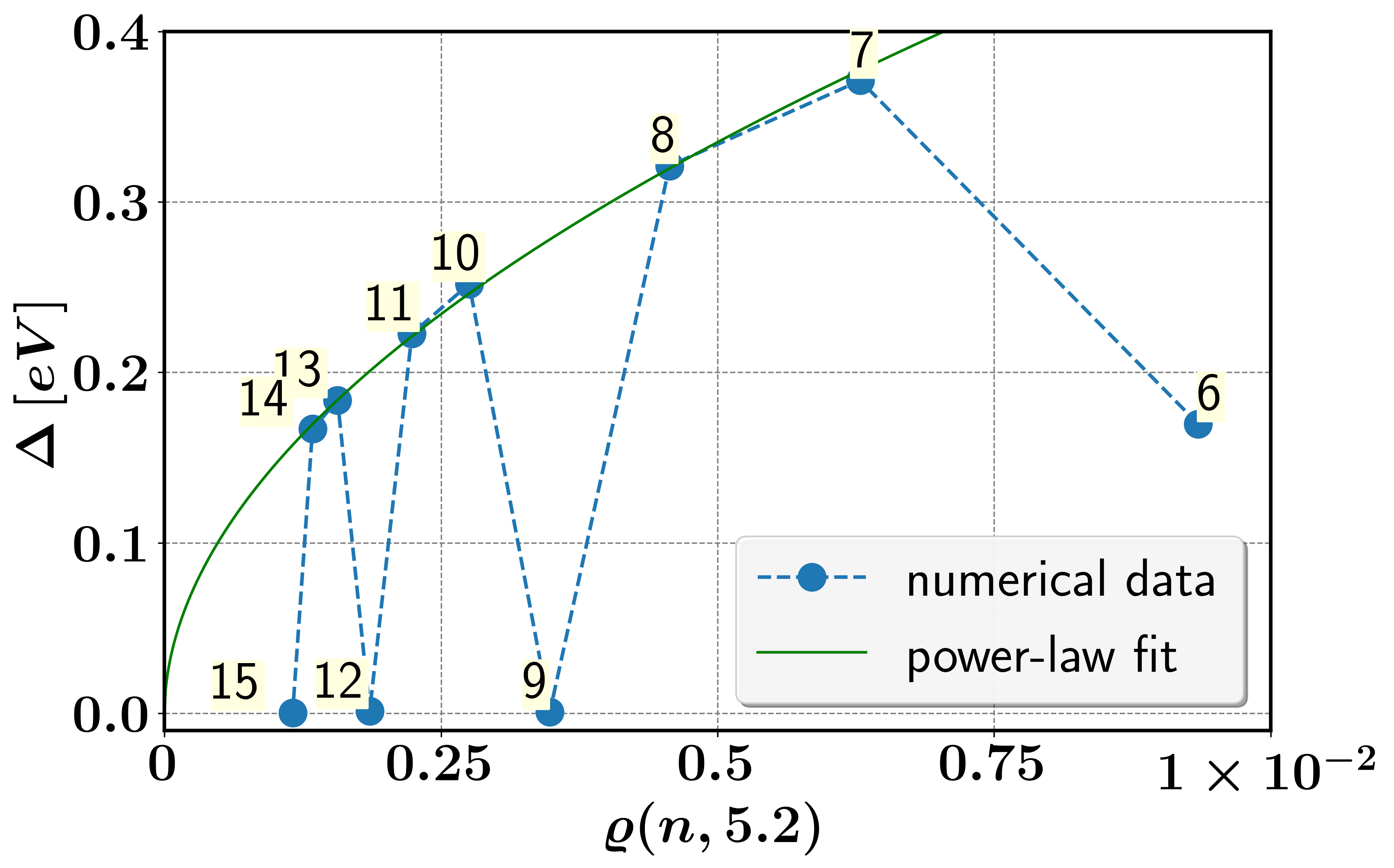}\\
\fl
c)	\includegraphics[scale=0.3]{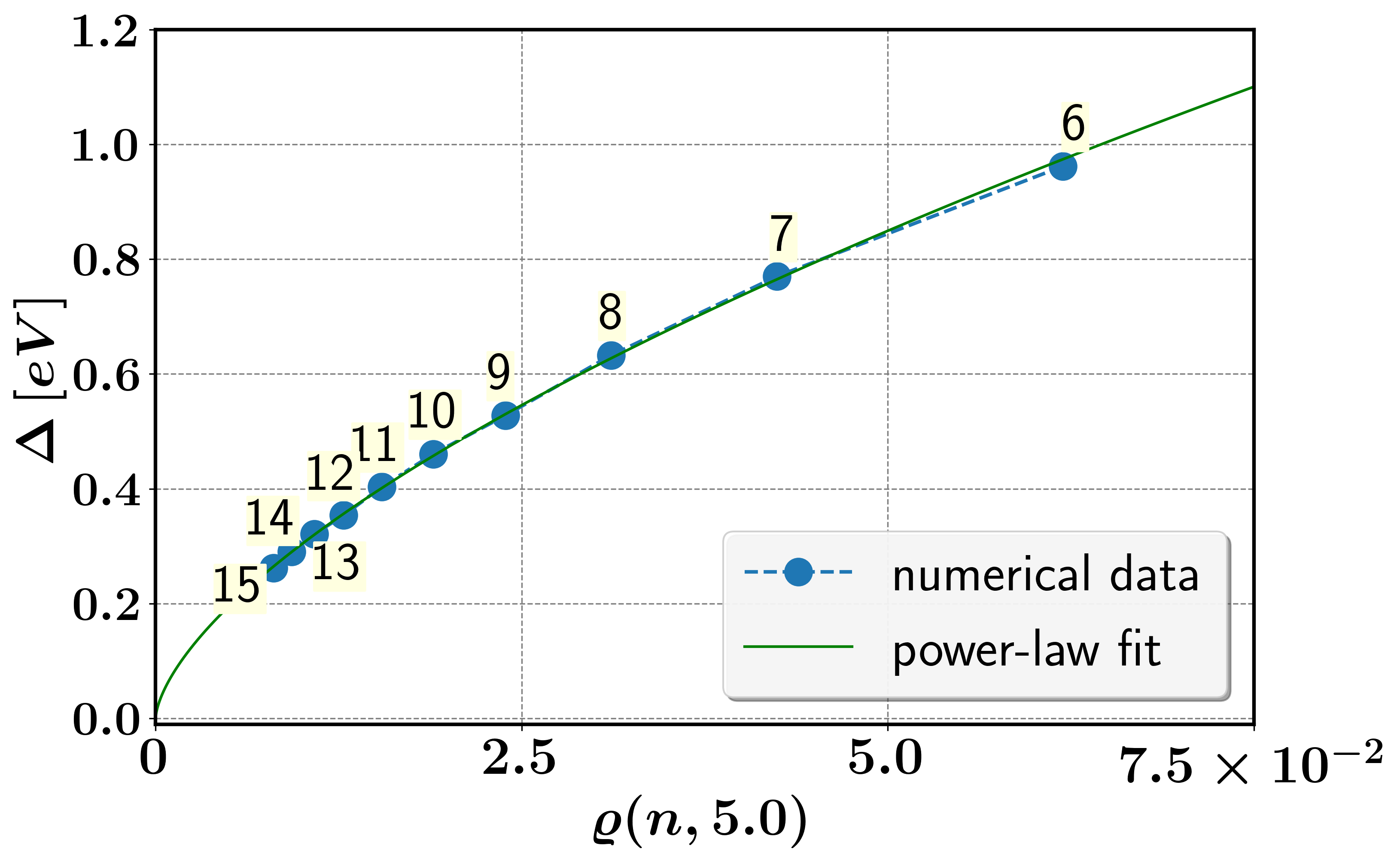}
\caption{Gap evolution as a function of $n$ for different hole radius. Panel a) is for $R=3$ and panel b) $R=5.2$, both corresponding to systems without dangling bonds. Panel c) corresponds to $R=5.0$ 
  which contains dangling bonds.  In cases a) and b), gapless systems occur with periodicity of $\Delta n=3$. Otherwise, gaps are given by power law envelops in the form of $\Delta= \Delta_{0} \varrho(n,R)^{p}$. The following fittings laws were obtained, a) $\Delta= 7.40\,\, \varrho(n,3.0)^{0.5797}, \overline{R}^{2}= 0.9997$. b) $\Delta=5.1732 \varrho(n,5.2)^{0.5166}, \overline{R}^{2}= 0.9977$ and c) $\Delta=5.750 \varrho(n,3.0)^{0.6384}, \overline{R}^{2}= 0.9997$ where $\overline{R}$ is the statistical coefficient of determination. \label{fig:5}}
\end{figure}

Let us know discuss the origin of the gaps. In Fig. \ref{fig:5} we present the gap sizes  for three distinct graphene series with hole radii of $R=3, 5, 5.2$ \AA, respectively. For each of these series, the size of the unit cell,$n$, is considered within the range of values  $n=6, \ldots, 15$, with the exception of the first series where $n=4, \ldots,15$. The initial observation of note in Fig. \ref{fig:5} is that, for the series featuring hole radii of $R=3$ and $R=5.2$ \AA, a periodicity with the property $\sim n \equiv 0 , (\mbox{mod} 3)$ emerges for $n$  sufficiently large values relative to the specific $R$ value at which the gap reaches zero. This phenomenon is particularly evident in Figure \ref{fig:3} c) and e), where the emergence of Dirac cones at the  $\Gamma$ points is clearly observed.	
 To understand the origin of the gaps, we rely on Fig. \ref{fig:5}, and excluding the semimetallic cases, it is obtained that in the three series the gap size, $\Delta$,  follows a power law as a function of the imbalance density $\sim \varrho(n, R)^p$, where $p$ is approximately $0.5166$, $0.5797$, and $0.6384$ for the respective series. These values, close to $p=1/2$, are comparable to the case where there is a density of impurities breaking the sublattice symmetry, and the imbalance density is given by $x_{\mbox{imp}}$. In a previous work \cite{Naumis2007}, it was found analytically that $|\Delta| \approx |t_{0}| \sqrt{6 x} \approx 6.85857 x^{0.5}$. The nearly similar exponents indicate that in the end, big holes enter as effective impurities and that the mechanism of underlying frustration due to sublattice imbalance is in play here \cite{Naumis2007}.

On other hand, we will now demonstrate that gap  periodicity arises due to the folding of the Dirac cones from the pristine graphene hexagonal Brillouin zone to the holey superlattice rectangular Brillouin zone (see Fig. \ref{fig:4} c) and from broken  symmetries. To do this, consider the original Dirac points on each of graphene's valley,
  \begin{equation}
      \boldsymbol{K}_{\xi}= \xi \frac{4 \pi}{3 \sqrt{3}a_{0}} (1,0), \,\, \xi= \pm 1
    \end{equation}
in the hexagonal Brillouin zone of graphene and $\xi$ is the valley  pseudospin index. On other hand,  the reciprocal vectors 
   \begin{equation}
       \boldsymbol{b}_{1}= \frac{2 \pi}{\sqrt{3} n a_{0}}(1,0), \hspace{1mm}
       \boldsymbol{b}_{2}= \frac{2 \pi}{3m a_{0}} (0,1)
   \end{equation}
   of holey graphene with UCHG$(n,m,R)$. It can be demonstrated that if $n= 3 l + \eta$ with $l \in \mathbb{Z}$ and $\eta=-1,0,1$, the following equation is obtained,
\begin{equation} \label{eq: folding-K-point}
			\boldsymbol{K}_{\xi}=\xi(2l+\eta) \boldsymbol{b}_{1} - \xi \frac{\eta}{3} \boldsymbol{b}_{1} = \boldsymbol{G}_{\xi (2l+ \eta),0}+ \boldsymbol{\kappa}_{\xi,\eta}; \,\, \, \boldsymbol{\kappa}_{\xi, \eta} \equiv - \frac{\xi \eta}{3} \boldsymbol{b}_{1}
   \end{equation} 

\begin{figure}[t]
		\fl
		a)\includegraphics[width=0.44\textwidth]{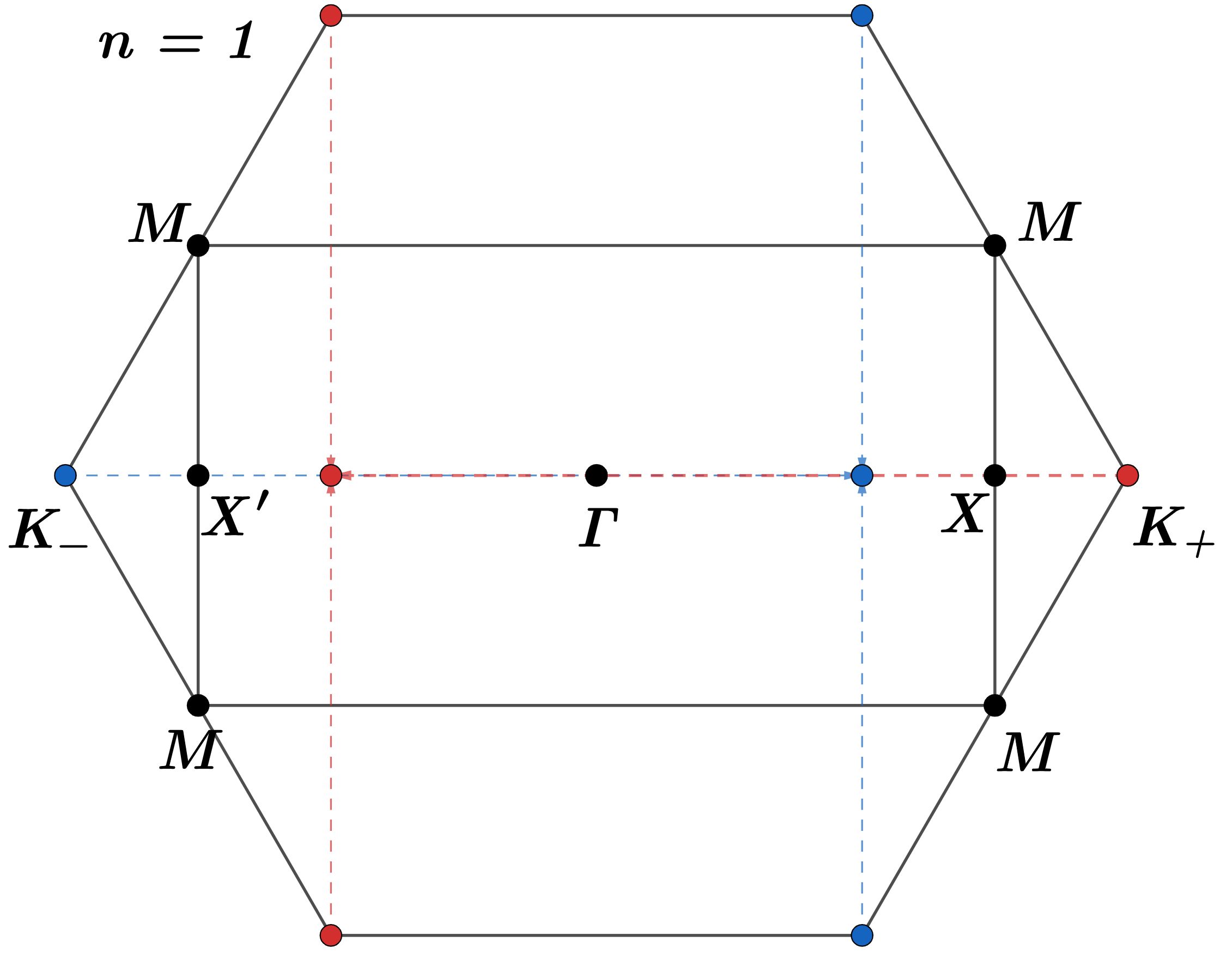}
		b) \includegraphics[width=0.44\textwidth]{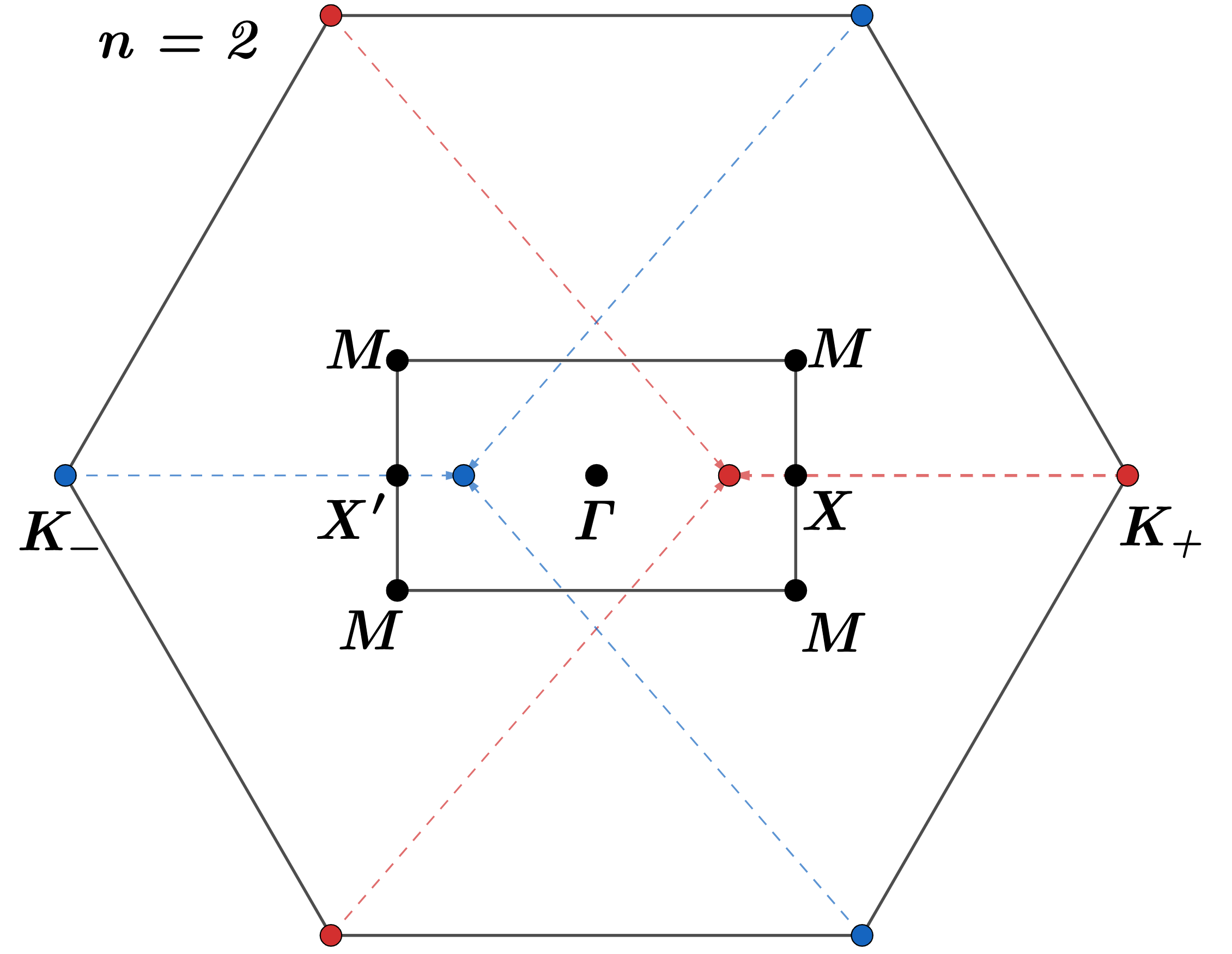}\\
	\fl	c) \includegraphics[width=0.44\textwidth]{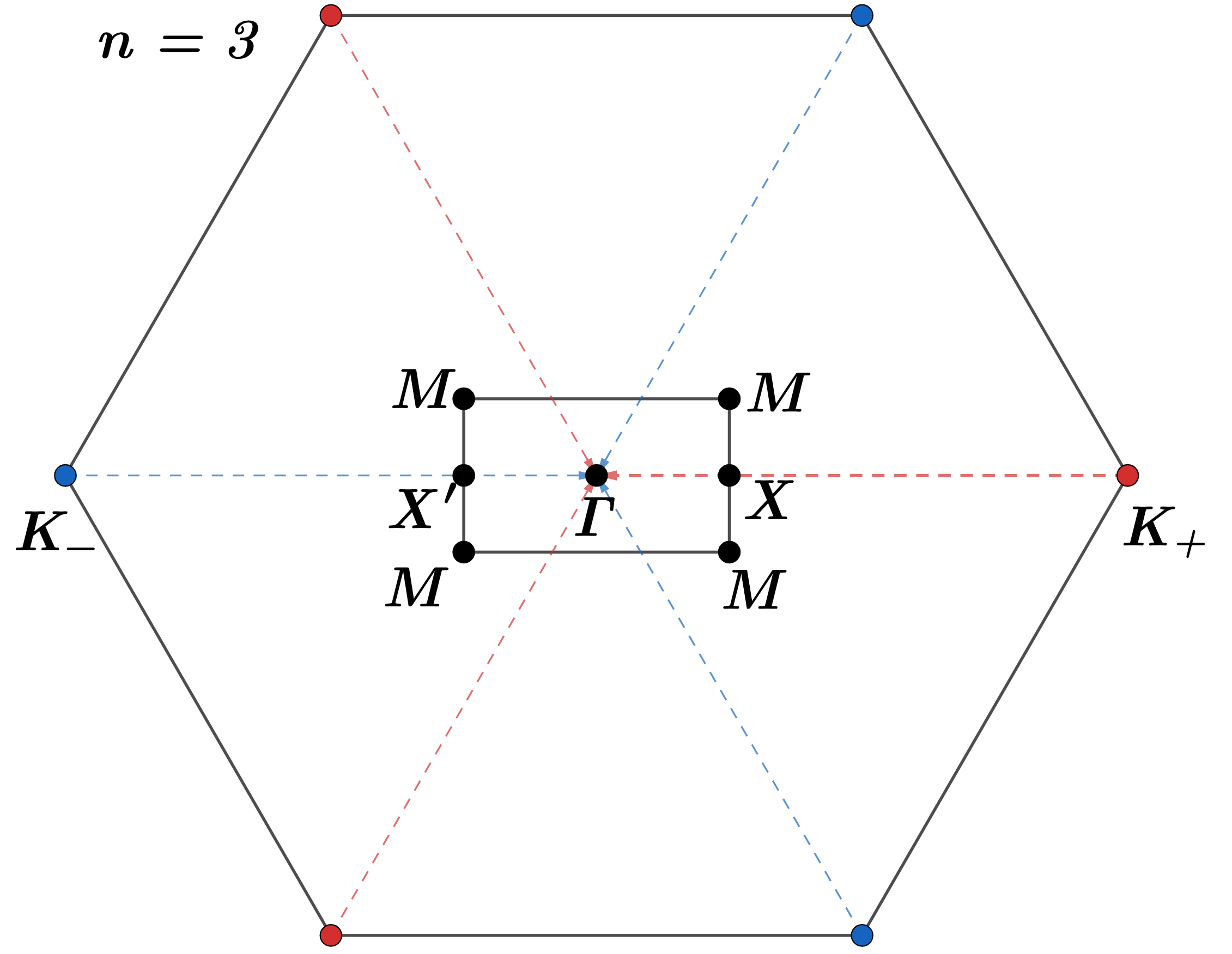}
		d) \includegraphics[width=0.44\textwidth]{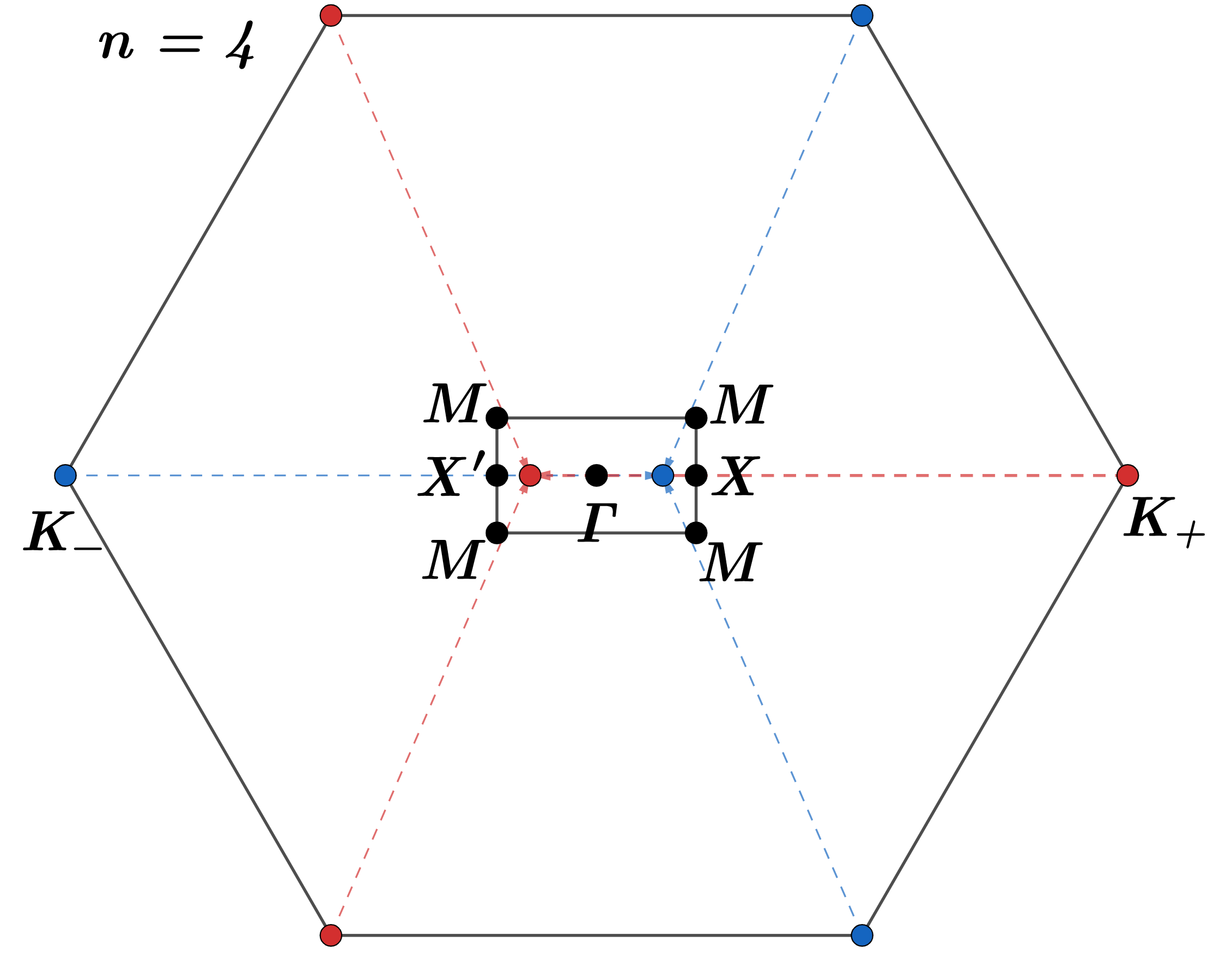}\\
		\fl
		e) \includegraphics[width=0.44\textwidth]{Fig-6.5.png}
		f) \includegraphics[width=0.44\textwidth]{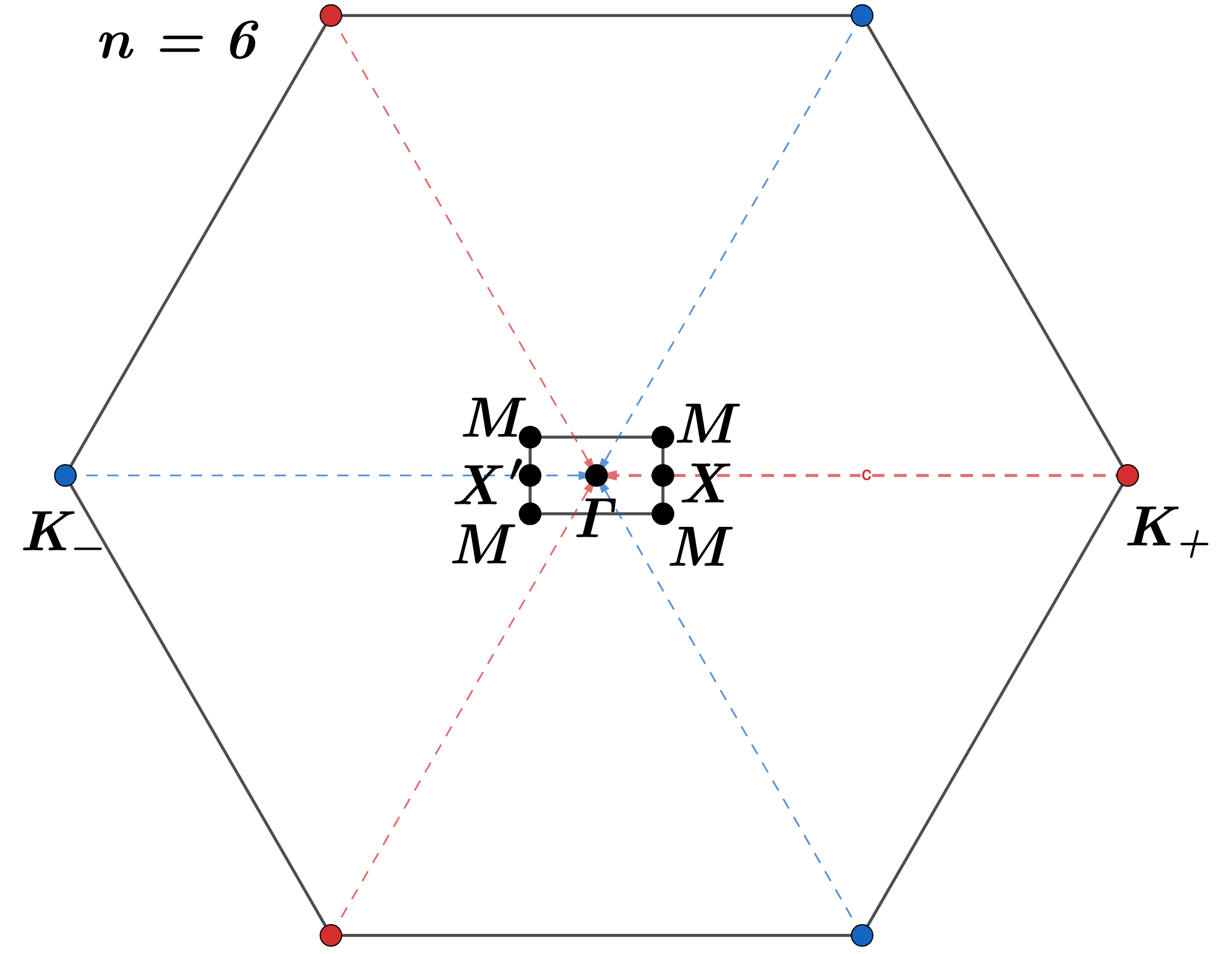}
		\caption{Brillouin zone of graphene and holey graphene with unit cell UCHG$(n,n,r)$ with $n=1, \ldots, 9$. The $K_{\pm}$ points are shown to be folded into $\pm \eta/3 \boldsymbol{b}_{1}$ for $\eta=-1,0,1$ and $n \equiv \eta $ (mod $3$) (see Eq. \ref{eq: folding-K-point}. In particular, for $n \equiv 0$ (mod $3$), the $K_{\pm}$ folds to the $\Gamma$ point, representing a valley degeneracy. \label{fig:supp-1}}
	\end{figure}

Therefore, $\boldsymbol{K}_{\xi}$ folds onto the point $\boldsymbol{\kappa}_{\xi,\eta}$, i.e., along $\Gamma$ to $X=\boldsymbol{b}_{1}/2$ or $\Gamma$ to $X'=-\boldsymbol{b}_{1}/2$ path  and its corresponding valley pseudospin $\xi$ and showing the periodicity of $\Delta n= 3$ through the $\eta$ index, and is independent of the size index $m$ or hole radii $R$.  Fig. \ref{fig:supp-1}) presents such folding sequence from $n=1$ to $n=6$.

It is noteworthy that when $\eta=0$, $K_{\pm}$ folds to the $\Gamma$ point, leading to valley degeneracy. Upon introducing holes in graphene (without dangling bonds), sublattice and inversion symmetries are broken, resulting in gaps at the Dirac points for $\eta= \pm 1$ \cite{ Wang2016, Kou2014, NISHIDATE2023122196, AGRAWAL2013102, Malterre_2011, Zhou2007}. However, $\Gamma$ is a point protected by other symmetries, ensuring the preservation of the two degenerate cones when $\eta=0$. To induce a gap at $\Gamma$, dangling bonds are necessary, which breaks bond symmetry \cite{Jia2016, YE201460}. 

Therefore, the imbalance density, the folding of the Brillouin zone, and symmetry breaking explain the behavior of the gap size shown in Fig. \ref{fig:5}.

\begin{figure}[t]
		\fl
		a)\includegraphics[scale=0.4]{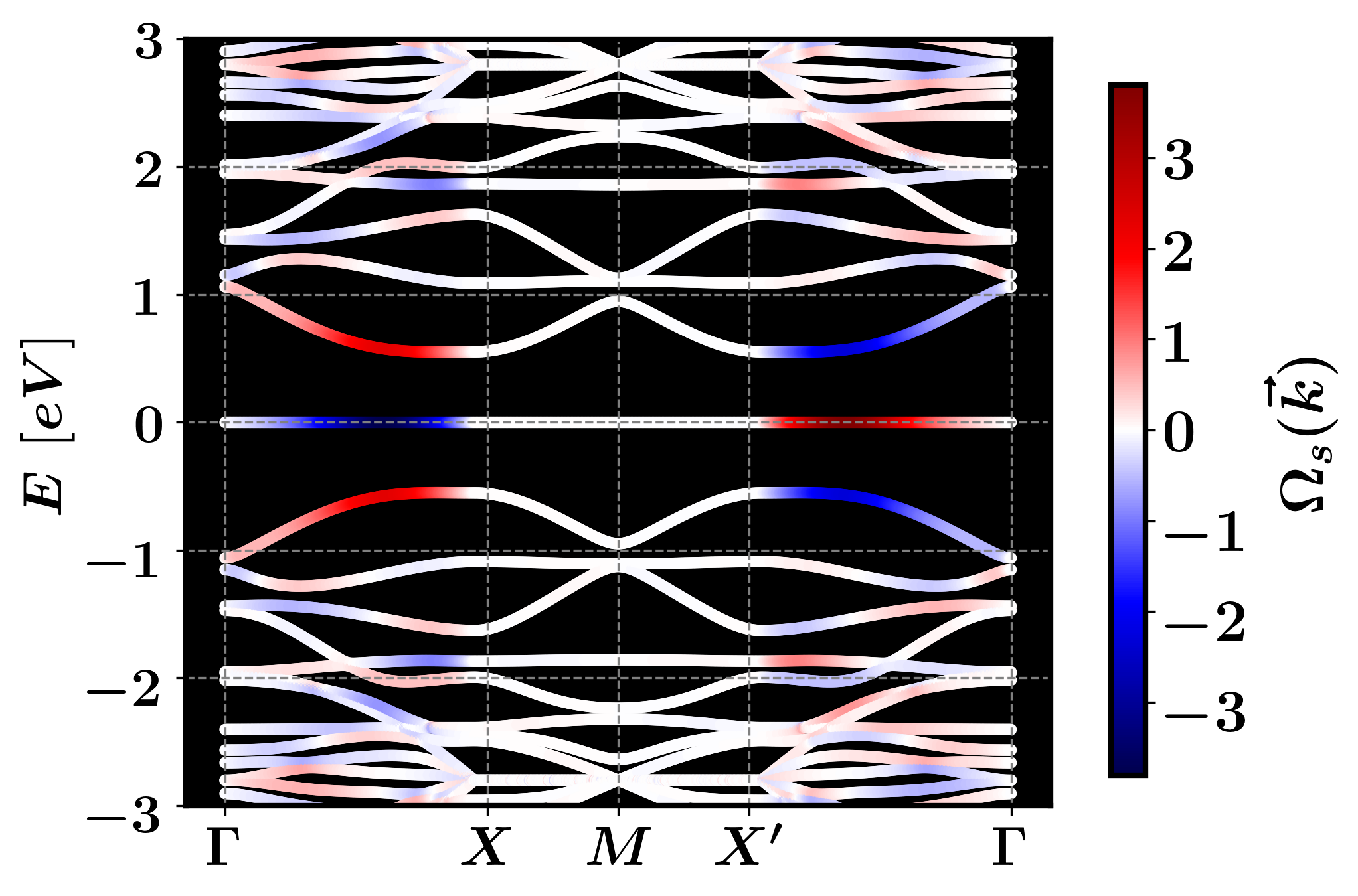}
		b) \includegraphics[scale=0.4]{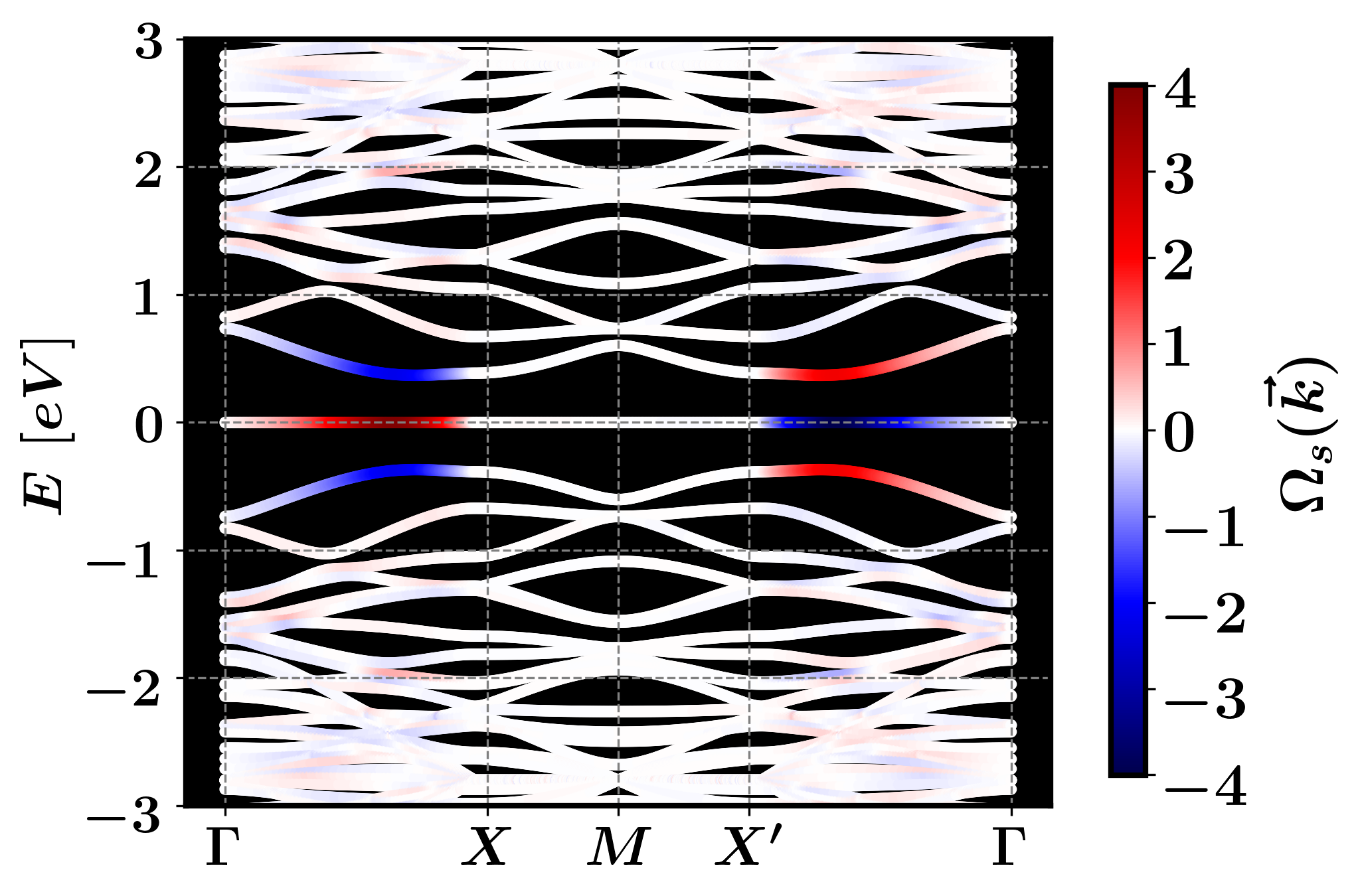}\\
		\fl
		c) \includegraphics[scale=0.4]{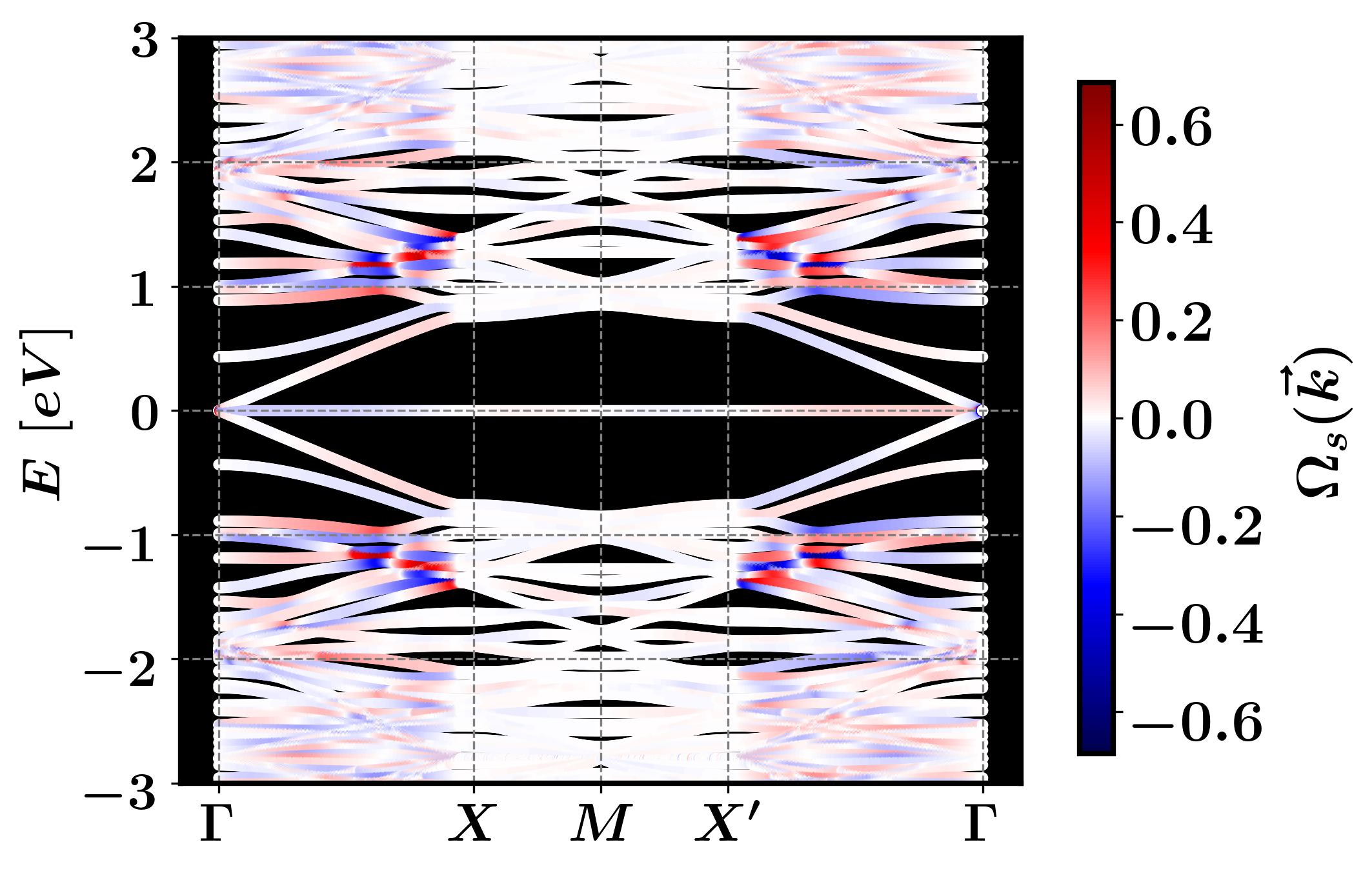}
		d) \includegraphics[scale=0.4]{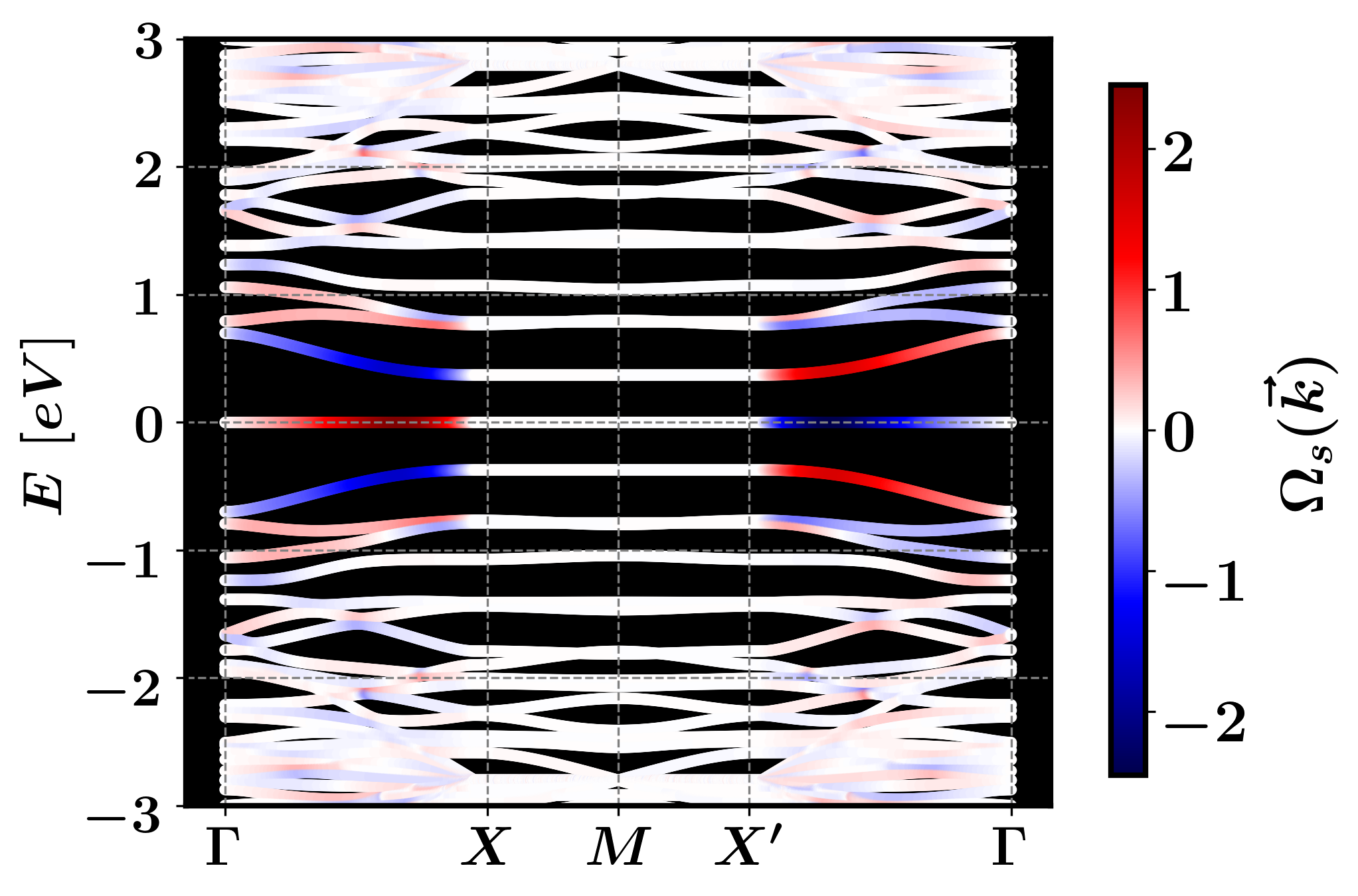}\\
		\fl
		e) \includegraphics[scale=0.4]{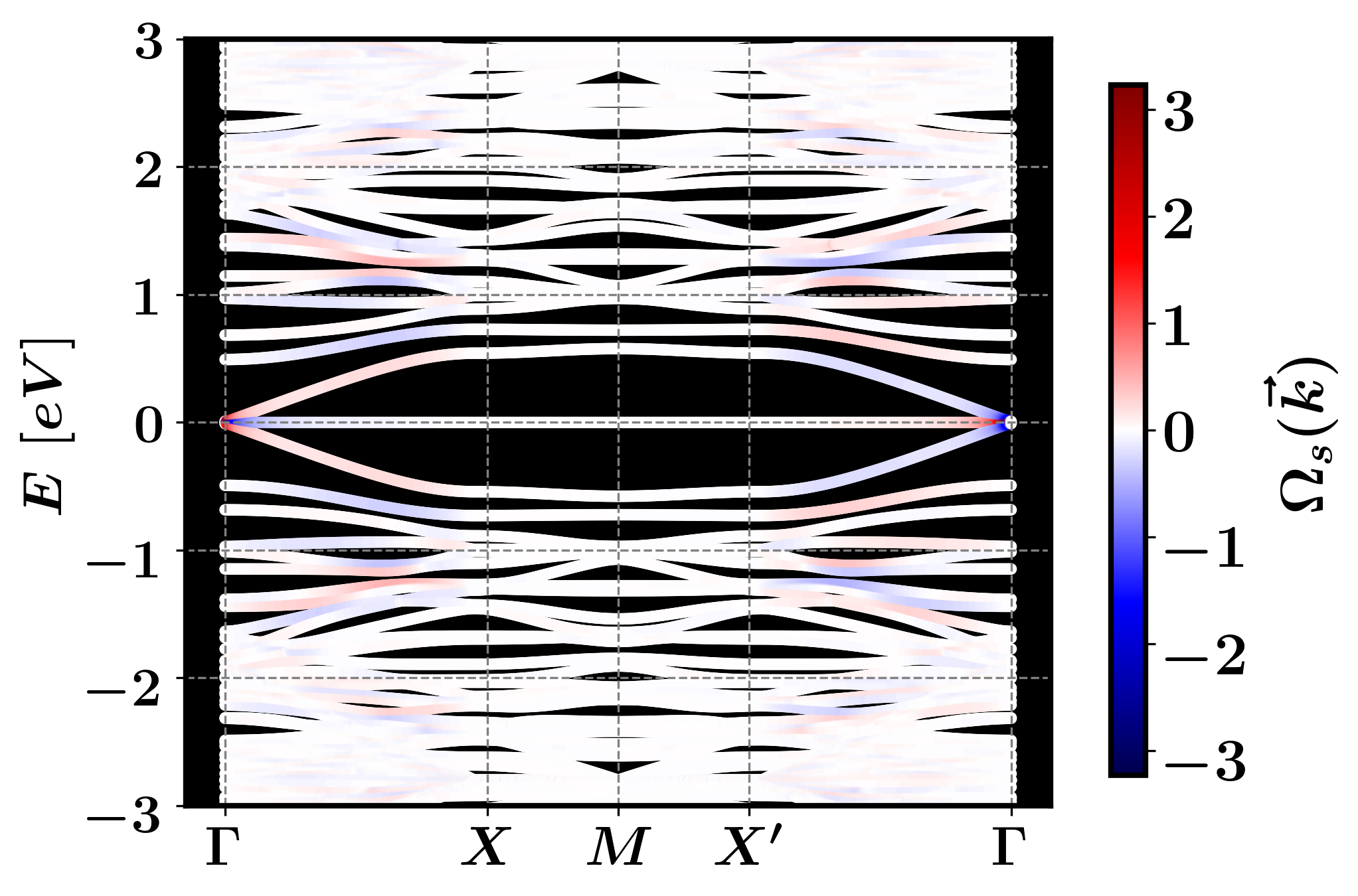}
		f) \includegraphics[scale=0.4]{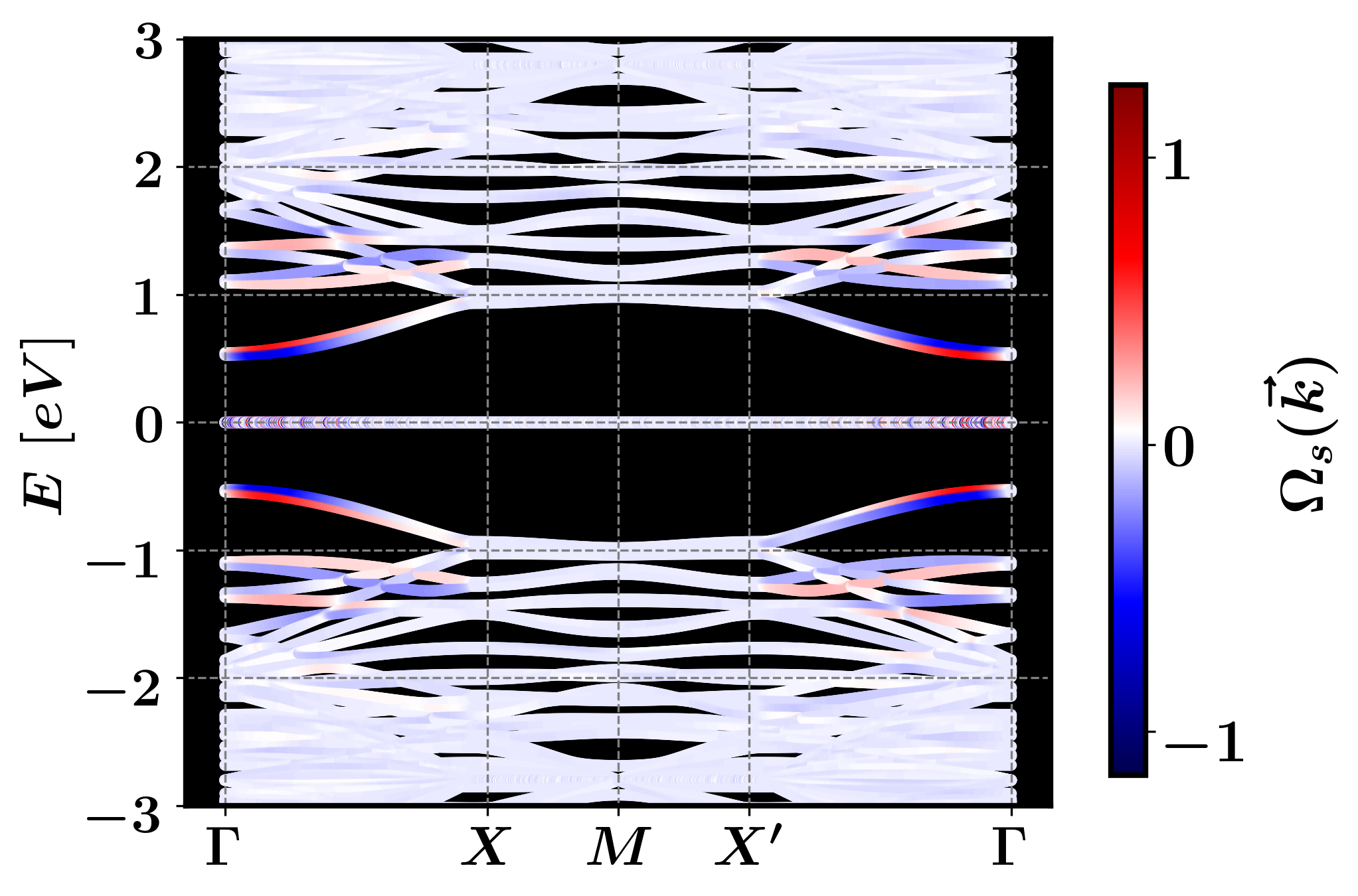}
		\caption{As discussed in Eq. \ref{eq:Berry curvature}, the symmetries broken induce a nonzero Berry curvature at the folding points $K_{\pm}$ and in the vicinity of $E=0$.   \label{fig:supp-4}}
	\end{figure}
  
Additionally, Fig. \ref{fig:supp-4} shows the Berry curvature for the unit cells shown in Fig. \ref{fig:2-1} as a function of $\boldsymbol{k}$ and the band $s$. The Berry curvature is denoted as $\Omega_{s}(\boldsymbol{k})$ and defined as,
\begin{eqnarray}\label{eq:Berry curvature}
\boldsymbol{\Omega}_{s}(\boldsymbol{k})= \nabla_{\boldsymbol{k}} \times \mathcal{A}_{s} (\boldsymbol{k}) \nonumber \\
\mathcal{A}_{s}(\boldsymbol{k})= i \braket{u_{s \boldsymbol{k}}| \nabla_{\boldsymbol{k}}| u_{s, \boldsymbol{k}}}
\end{eqnarray}

where $\mathcal{A}_{s}(\boldsymbol{k})$ is the Berry connection and $u_{s, \boldsymbol{k}}$ are the periodic functions in the Bloch states that solve the Schrödinger equation. Figure  Fig. \ref{fig:supp-4} illustrates how the symmetry breaking induces a nonzero Berry curvature at the folding points $K_{\pm}$ and in the vicinity of $E=0$.
		
Finally, near the high-symmetry points $\boldsymbol{\kappa}_{\eta \xi}$ of the reciprocal folded lattice and without dangling bond  three symmetric bands around $E=0$ are formed, one for valence, the flat band, and the conduction band, denoted by  $s=-1, 0, 1$, respectively. Folding the Dirac points at $\boldsymbol{\kappa}_{\xi,\eta}$ would, in principle, yield two bands with linear dispersion and a flat band at $E=0$. Therefore, an appropriate Hamiltonian for this description is the $\alpha-\mathcal{T}_{3}$ model \cite{Mojarro2020}, defined by

\begin{equation} \label{eq:alpha-T3 general model}
\hat{H}_{0}(\alpha)= \hbar v_{F} \boldsymbol{\hat{S}}_{\xi}(\alpha) 
\end{equation}
where  $\boldsymbol{\hat{S}_{\xi}}(\alpha)= \left( \xi \hat{S}_{x}(\alpha), \hat{S}_{y}(\alpha)\right)$ with $0 \leq \alpha \leq 1$ a parameter, and 
$\hat{S}_{x}(\alpha), \hat{S}_{y}(\alpha)$, and $v_{F}$ are the  pseudospin operators and Fermi velocity, respectively;  given by 
\begin{eqnarray} \label{eq: pseudospin operators 1}
v_{F}= \mathfrak{c}(n,R) \frac{t_{0}a_{0}}{\hbar}\nonumber \\
\hat{S}_{x}(\alpha)= \left( \begin{array}{ccc}
   0& C_{\alpha} &  0 \\
   C_{\alpha}  & 0 & S_{\alpha}\\
   0 & S_{\alpha } & 0
\end{array} \right) \nonumber \\
\hat{S}_{y}(\alpha)= \left( \begin{array}{ccc}
   0& -iC_{\alpha} &  0 \\
   iC_{\alpha}  & 0 & -iS_{\alpha}\\
   0 & i S_{\alpha } & 0
\end{array} \right) 
\end{eqnarray}
with $C_{\alpha}= 1/\sqrt{1+ \alpha^{2}}$, $S_{\alpha}= \alpha/ \sqrt{1+ \alpha^{2}}$ and $\mathfrak{c}(n,R)$ a constant that depends on unit cell size $n$ and hole size $R$. Such hamiltonian is the most simple one that has flat bands and under electromagnetic radiation, behaves as a two- or three-level Rabi system with clear optical signatures of flat bands \cite{Mojarro2020}.

As discussed previously, breaking inversion and sublattice symmetry results in the opening of a gap. Therefore, the low-energy Hamiltonian around $\boldsymbol{\kappa}_{\xi,\eta}$ will exhibit an additional mass-like term to the $\alpha-\mathcal{T}_{3}$ Hamiltonian, thus the low-energy hamiltonian is
\begin{equation} \label{eq:low-energy-hamiltonian}
\mathcal{H}_{\eta \xi}(\boldsymbol{k})= \hbar v_{F} \boldsymbol{\hat{S}}_{\xi}(|\xi|) \cdot \boldsymbol{k}+ M_{\tau}(\boldsymbol{k}) \Sigma_{z}
\end{equation} 
where $\Sigma_{z}$ is a pseudospin operator defined as
\begin{equation} \label{eq:sigma_z pseudospin}
    \Sigma_{z}=\left( \begin{array}{ccc}
         1&0&0  \\
         0&0&0\\
         0&0&-1
    \end{array}\right)
\end{equation}

and $M_{\tau}(\boldsymbol{k})$ is the mass-like term that depends on $\boldsymbol{k}$ and pseudospin index $\tau= \xi \eta= -1,0,1 $. It can be constructed from the expansion of energies for the bands $s$, given by

\begin{eqnarray}\label{eq:energies-expansion} 
\fl
E_{s}(\boldsymbol{\kappa}_{\xi,\eta}+\boldsymbol{q})& \approx \braket{u_{s, \boldsymbol{\kappa}_{\xi,\eta}}|\hat{\mathcal{H}_{T}}(\boldsymbol{\kappa}_{\xi,\eta})| u_{s, \boldsymbol{\kappa}_{\xi,\eta}}} \nonumber\\
& + \braket{u_{s, \boldsymbol{\kappa}_{\xi,\eta}}\left|\left.\boldsymbol{q} \cdot \left( \nabla_{\boldsymbol{k}} \hat{\mathcal{H}}_{T}(\boldsymbol{k})\right)\right|_{\boldsymbol{k}= \boldsymbol{\kappa}_{\xi,\eta}} \right|u_{s, \boldsymbol{\kappa}_{\xi,\eta}}}+ \mathcal{O}(|\boldsymbol{q}|^{2}) \nonumber\\
&=s\left( |E_{1}(\boldsymbol{\kappa}_{\xi,\eta})|+ \mathfrak{a}_{2}(n,R) q_{x}+ \mathcal{O}(|\boldsymbol{q}|^{2})\right).
\end{eqnarray}
where $|E_{1}(\boldsymbol{\kappa}_{\xi,\eta})|, \,\mathfrak{a}_{2}(n,R)$ can be numerically determined. Then $M_{\tau}(\boldsymbol{k})$ is expressed as,
\begin{equation} \label{eq:mass-like-term-definition}
 M_{\tau}(\boldsymbol{k})= |E_{1}(\boldsymbol{\kappa}_{\xi,\eta})|- \tau |\mathfrak{a}_{2}(n,R)| k_{x} 
\end{equation}

\begin{figure}[t]
\fl
		a)\includegraphics[scale=0.38]{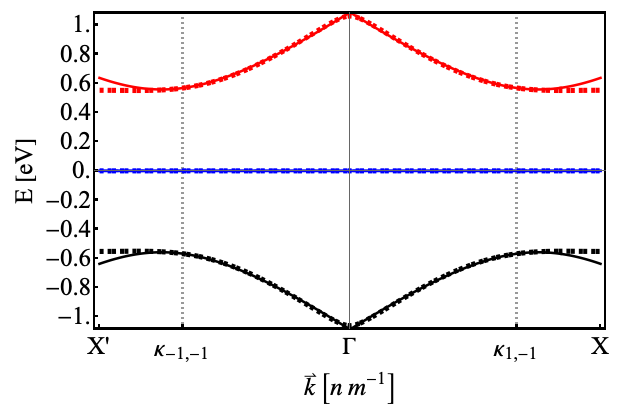}
		b) \includegraphics[scale=0.38]{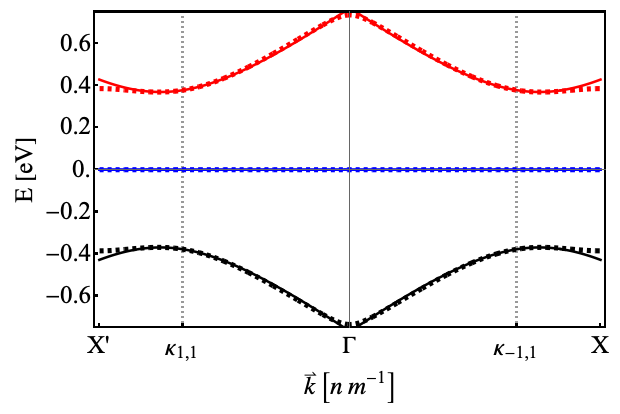}\\
\fl
		c) \includegraphics[scale=0.38]{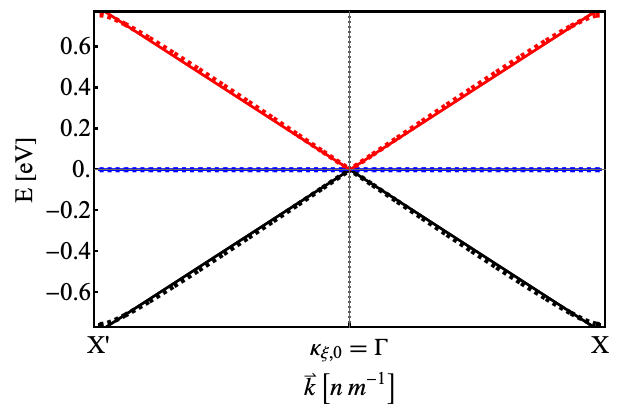}
		d) \includegraphics[scale=0.38]{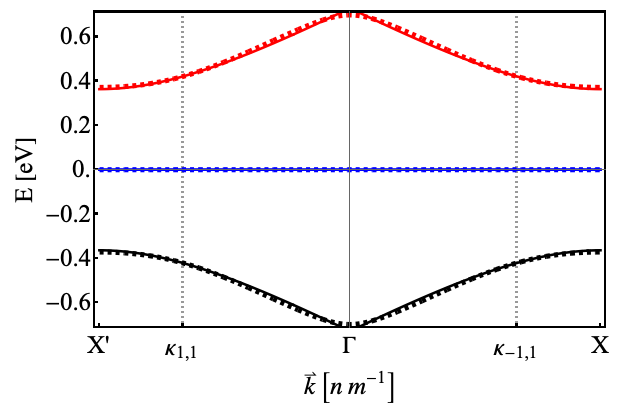}\\
\fl
		e) \includegraphics[scale=0.38]{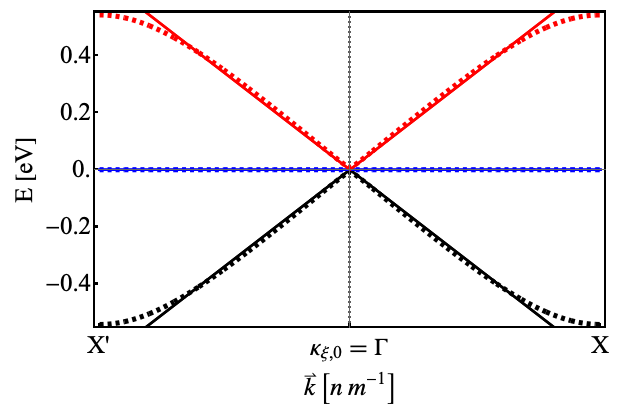}
		\caption{ The electronic bands obtained from the Hamiltonian $\hat{\mathcal{H}}_{T}$ (dashed black, blue, and red lines) are compared with the low-energy approximation using the Hamiltonian $\mathcal{H}_{\xi, \eta}$ (solid black, blue, and red lines) for a) UCHG$(5,5,3)$, $\mathfrak{c}(5,3)=\frac{6}{5}$, $|E_{1}(\boldsymbol{\kappa}_{\xi,\eta})|\approx 0.5680$ eV, $|\mathfrak{a}_{2}(5,3)| \approx 0.0903$ eV nm, b) UCHG$(7,7,3)$, $\mathfrak{c}(7,3)=\frac{6}{5}$, $|E_{1}(\boldsymbol{\kappa}_{\xi,\eta})|\approx 0.3780$ eV, $|\mathfrak{a}_{2}(7,3)| \approx 0.1031$ eV nm, c) UCHG$(9,9,3)$, $\mathfrak{c}(9,3)=\frac{7}{5}$, $|E_{1}(\boldsymbol{\kappa}_{\xi,\eta})|=0$ eV, $|\mathfrak{a}_{2}(9,3)|=0$ eV nm, d) UCHG$(7,7,5.2)$, $\mathfrak{c}(7,5.2)=\frac{3}{4}$, $|E_{1}(\boldsymbol{\kappa}_{\xi,\eta})|\approx 0.4209$ eV, $|\mathfrak{a}_{2}(7,5.2)| \approx 0.1739$ eV nm and f) UCHG$(9,9,5.2)$, $\mathfrak{c}(9,5.2)=\frac{6}{5}$, $|E_{1}(\boldsymbol{\kappa}_{\xi,\eta})|=0$ eV, $|\mathfrak{a}_{2}(9,5.2)| =0$ eV nm. See Eqs. \ref{eq:alpha-T3 general model}- \ref{eq:mass-like-term-definition}. An excellent agreement between the low energy Hamiltonian and the tight-binding Hamiltonian is seen \label{fig:supp-5}}
	\end{figure}

In Fig. \ref{fig:supp-5}, the electronic bands obtained from the Hamiltonian $\hat{\mathcal{H}}_{T}$ (dashed black, blue and red lines) and the low-energy approximation with the Hamiltonian $\mathcal{H}_{\xi, \eta}$ (Eq. \ref{eq:low-energy-hamiltonian}) (solid black, blue and red lines) are depicted. In general a excellent agreement is obtained.

	\section{Concluding remarks \label{sec:conclusions}}

In this paper, we investigated the formation of flat bands in two-dimensional periodic holey graphene (2D HG). Our findings reveal that these flat band states exhibit significantly higher localization when compared to other states, particularly in the zigzag edge regions surrounding the hole. This highlights a connection between flat band formation and Compact Localized States (CLS), as investigated in prior works \cite{Naumis1994,Naumis2002, Espinosa-Champo_2024}. Moreover, we establish that sublattice site imbalance, achieved through the breaking of path-exchange symmetry \cite{Jun2023, Espinosa-Champo_2024}, induces the formation of flat bands, while the breaking of sublattice and inversion symmetry leads to the creation of energy gaps.

Furthermore, we have discussed the influence of unit cell size and hole dimensions on the gap size and density of states (DOS) at $E=0$, through the imbalance density $\varrho(n,R)$. Additionally, we have explored the existence of semimetallicity in 2D HG with a periodicity of $\Delta n=3$, arising from the folding of $K_{\pm}$
points to the $\Gamma$ point, protected by bond symmetry, except when introducing dangling bonds that break this symmetry, resulting in a gap. We demonstrated that the breaking of the discussed symmetries (inversion, sublattice, and bond) generates a non-zero Berry curvature.

Finally, we establish a continuous Hamiltonian model for three bands in cases where the hole radius $R\neq 0$ and there are no dangling bonds. This model consists of an $\alpha-\mathcal{T}_{3}$
type Hamiltonian and a mass-like term that opens the gap at $\boldsymbol{\kappa}_{\xi \eta}$ points.

Our work gives a protocol that allows to obtain flat bands at will and thus gives a possible new 2D material to obtain highly correlated quantum phases without twists. 
 
	\section*{Acknowledgements}
	This work was supported by  CONAHCyT project 1564464 and UNAM DGAPA project IN101924. Abdiel de Jesús Espinosa-Champo is supported by a CONAHCyT PhD fellowship (No. CVU 1007044).  
	The authors acknowledge and express gratitude to Carlos Ernesto L\'opez Natar\'en from Secretaria T\'ecnica de C\'omputo y Telecomunicaciones for  his valuable support.


 \newpage

	\section*{References}
	\bibliographystyle{iopart-num}

\begin{thebibliography}{10}
	\expandafter\ifx\csname url\endcsname\relax
	\def\url#1{{\tt #1}}\fi
	\expandafter\ifx\csname urlprefix\endcsname\relax\def\urlprefix{URL }\fi
	\providecommand{\eprint}[2][]{\url{#2}}
	
	\bibitem{QiuWenXuan}
	Qiu W~X, Li S, Gao J~H, Zhou Y and Zhang F~C 2016 {\em Phys. Rev. B\/} {\bf
		94}(24) 241409
	\urlprefix\url{https://link.aps.org/doi/10.1103/PhysRevB.94.241409}
	
	\bibitem{Drost_Ojanen_Harju_Liljeroth_2017}
	Drost R, Ojanen T, Harju A and Liljeroth P 2017 {\em Nature Physics\/} {\bf 13}
	668–671 ISSN 1745-2481
	
	\bibitem{Abilio1999}
	Abilio C~C, Butaud P, Fournier T, Pannetier B, Vidal J, Tedesco S and Dalzotto
	B 1999 {\em Phys. Rev. Lett.\/} {\bf 83}(24) 5102--5105
	\urlprefix\url{https://link.aps.org/doi/10.1103/PhysRevLett.83.5102}
	
	\bibitem{HidekiOzawa2015}
	Taie S, Ozawa H, Ichinose T, Nishio T, Nakajima S and Takahashi Y 2015 {\em
		Science Advances\/} {\bf 1} e1500854 (\textit{Preprint}
	\eprint{https://www.science.org/doi/pdf/10.1126/sciadv.1500854})
	\urlprefix\url{https://www.science.org/doi/abs/10.1126/sciadv.1500854}
	
	\bibitem{Nakata2012}
	Nakata Y, Okada T, Nakanishi T and Kitano M 2012 {\em Phys. Rev. B\/} {\bf
		85}(20) 205128
	\urlprefix\url{https://link.aps.org/doi/10.1103/PhysRevB.85.205128}
	
	\bibitem{He_Mao_Cai_Zhang_Li_Yuan_Zhu_Wang_2021}
	He Y, Mao R, Cai H, Zhang J~X, Li Y, Yuan L, Zhu S~Y and Wang D~W 2021 {\em
		Physical Review Letters\/} {\bf 126} 103601 ISSN 0031-9007, 1079-7114
	
	\bibitem{Cao2018}
	Cao Y, Fatemi V, Fang S, Watanabe K, Taniguchi T, Kaxiras E and Jarillo-Herrero
	P 2018 {\em Nature\/} {\bf 556} 43--50
	\urlprefix\url{https://doi.org/10.1038/nature26160}
	
	\bibitem{Leonardo2021}
	Naumis G~G, Navarro-Labastida L~A, Aguilar-M\'endez E and Espinosa-Champo A
	2021 {\em Phys. Rev. B\/} {\bf 103}(24) 245418
	\urlprefix\url{https://link.aps.org/doi/10.1103/PhysRevB.103.245418}
	
	\bibitem{Leonardo2022}
	Navarro-Labastida L~A, Espinosa-Champo A, Aguilar-Mendez E and Naumis G~G 2022
	{\em Phys. Rev. B\/} {\bf 105}(11) 115434
	\urlprefix\url{https://link.aps.org/doi/10.1103/PhysRevB.105.115434}
	
	\bibitem{Espinosa-Champo_2024}
	de~Jesús Espinosa-Champo A and Naumis G~G 2023 {\em Journal of Physics:
		Condensed Matter\/} {\bf 36} 015502
	\urlprefix\url{https://dx.doi.org/10.1088/1361-648X/acfbd1}
	
	\bibitem{Bergholtz_Liu_2013}
	Bergholtz E~J and Liu Z 2013 {\em International Journal of Modern Physics B\/}
	{\bf 27} 1330017 ISSN 0217-9792, 1793-6578
	
	\bibitem{Nguyen2018}
	Nguyen H~S, Dubois F, Deschamps T, Cueff S, Pardon A, Leclercq J~L, Seassal C,
	Letartre X and Viktorovitch P 2018 {\em Phys. Rev. Lett.\/} {\bf 120}(6)
	066102
	\urlprefix\url{https://link.aps.org/doi/10.1103/PhysRevLett.120.066102}
	
	\bibitem{DENG2003412}
	Deng S, Simon A and Köhler J 2003 {\em Journal of Solid State Chemistry\/}
	{\bf 176} 412--416 ISSN 0022-4596 special issue on The Impact of Theoretical
	Methods on Solid-State Chemistry
	\urlprefix\url{https://www.sciencedirect.com/science/article/pii/S0022459603002391}
	
	\bibitem{Mielke_Tasaki_1993}
	Mielke A and Tasaki H 1993 {\em Communications in Mathematical Physics\/} {\bf
		158} 341–371 ISSN 1432-0916
	
	\bibitem{Tasaki_1998}
	Tasaki H 1998 {\em Progress of Theoretical Physics\/} {\bf 99} 489–548 ISSN
	0033-068X
	
	\bibitem{AokiHideo2020}
	Aoki H 2020 {\em Journal of Superconductivity and Novel Magnetism\/} {\bf 33}
	2341--2346 \urlprefix\url{https://doi.org/10.1007/s10948-020-05474-6}
	
	\bibitem{WuCongjun2007}
	Wu C, Bergman D, Balents L and Das~Sarma S 2007 {\em Phys. Rev. Lett.\/} {\bf
		99}(7) 070401
	\urlprefix\url{https://link.aps.org/doi/10.1103/PhysRevLett.99.070401}
	
	\bibitem{Jaworowski_2018}
	Jaworowski B, Güçlü A~D, Kaczmarkiewicz P, Kupczyński M, Potasz P and Wójs
	A 2018 {\em New Journal of Physics\/} {\bf 20} 063023
	\urlprefix\url{https://dx.doi.org/10.1088/1367-2630/aac690}
	
	\bibitem{Leonardo2023}
	Navarro-Labastida L~A and Naumis G~G 2023 {\em Phys. Rev. B\/} {\bf 107}(15)
	155428 \urlprefix\url{https://link.aps.org/doi/10.1103/PhysRevB.107.155428}
	
	\bibitem{Leonardo2023RevMex}
	Navarro~Labastida L~A and G~Naumis G 2023 {\em Revista Mexicana de Física\/}
	{\bf 69} 041602 1–
	\urlprefix\url{https://rmf.smf.mx/ojs/index.php/rmf/article/view/6795}
	
	\bibitem{RomanTaboada2017}
	Roman-Taboada P and Naumis G~G 2017 {\em Phys. Rev. B\/} {\bf 95}(11) 115440
	\urlprefix\url{https://link.aps.org/doi/10.1103/PhysRevB.95.115440}
	
	\bibitem{RomanTaboada2017b}
	Roman-Taboada P and Naumis G~G 2017 {\em Phys. Rev. B\/} {\bf 96}(15) 155435
	\urlprefix\url{https://link.aps.org/doi/10.1103/PhysRevB.96.155435}
	
	\bibitem{RomanTaboada_2017JPC}
	Roman-Taboada P and Naumis G~G 2017 {\em Journal of Physics Communications\/}
	{\bf 1} 055023 \urlprefix\url{https://dx.doi.org/10.1088/2399-6528/aa98fd}
	
	\bibitem{Mao2020}
	Mao J, Milovanovi{\'c} S~P, An{\dj}elkovi{\'c} M, Lai X, Cao Y, Watanabe K,
	Taniguchi T, Covaci L, Peeters F~M, Geim A~K, Jiang Y and Andrei E~Y 2020
	{\em Nature\/} {\bf 584} 215--220
	\urlprefix\url{https://doi.org/10.1038/s41586-020-2567-3}
	
	\bibitem{Manesco_2021}
	Manesco A~L~R and Lado J~L 2021 {\em 2D Materials\/} {\bf 8} 035057
	\urlprefix\url{https://dx.doi.org/10.1088/2053-1583/ac0b48}
	
	\bibitem{Manesco_2021_2}
	Manesco A~L~R, Lado J~L, Ribeiro E~V~S, Weber G and Jr D~R 2020 {\em 2D
		Materials\/} {\bf 8} 015011
	\urlprefix\url{https://dx.doi.org/10.1088/2053-1583/abbc5f}
	
	\bibitem{Milanovic2020}
	Milovanović S~P, An\dj{}elković M, Covaci L and Peeters F~M 2020 {\em Phys.
		Rev. B\/} {\bf 102}(24) 245427
	\urlprefix\url{https://link.aps.org/doi/10.1103/PhysRevB.102.245427}
	
	\bibitem{Sandler2023}
	Mahmud M~T, Zhai D and Sandler N 2023 {\em Nano Letters\/} {\bf 23} 7725--7732
	pMID: 37578461 (\textit{Preprint}
	\eprint{https://doi.org/10.1021/acs.nanolett.3c02513})
	\urlprefix\url{https://doi.org/10.1021/acs.nanolett.3c02513}
	
	\bibitem{Elias2023}
	Andrade E, L\'opez-Ur\'{\i}as F and Naumis G~G 2023 {\em Phys. Rev. B\/} {\bf
		107}(23) 235143
	\urlprefix\url{https://link.aps.org/doi/10.1103/PhysRevB.107.235143}
	
	\bibitem{Guinea2017}
	Gonzalez-Arraga L~A, Lado J~L, Guinea F and San-Jose P 2017 {\em Phys. Rev.
		Lett.\/} {\bf 119}(10) 107201
	\urlprefix\url{https://link.aps.org/doi/10.1103/PhysRevLett.119.107201}
	
	\bibitem{Carr2018}
	Carr S, Fang S, Jarillo-Herrero P and Kaxiras E 2018 {\em Phys. Rev. B\/} {\bf
		98}(8) 085144
	\urlprefix\url{https://link.aps.org/doi/10.1103/PhysRevB.98.085144}
	
	\bibitem{Yndurain2019}
	Yndurain F 2019 {\em Phys. Rev. B\/} {\bf 99}(4) 045423
	\urlprefix\url{https://link.aps.org/doi/10.1103/PhysRevB.99.045423}
	
	\bibitem{Wu2021Pressure}
	Wu Z, Kuang X, Zhan Z and Yuan S 2021 {\em Phys. Rev. B\/} {\bf 104}(20) 205104
	\urlprefix\url{https://link.aps.org/doi/10.1103/PhysRevB.104.205104}
	
	\bibitem{Francisco2023}
	S\'anchez-Ochoa F, Rubio-Ponce A and L\'opez-Ur\'{\i}as F 2023 {\em Phys. Rev.
		B\/} {\bf 107}(4) 045414
	\urlprefix\url{https://link.aps.org/doi/10.1103/PhysRevB.107.045414}
	
	\bibitem{You2019}
	You J~Y, Gu B and Su G 2019 {\em Scientific Reports\/} {\bf 9} 20116 ISSN
	2045-2322 \urlprefix\url{https://doi.org/10.1038/s41598-019-56738-8}
	
	\bibitem{Sedelnikova2019}
	Sedelnikova O~V, Stolyarova S~G, Chuvilin A~L, Okotrub A~V and Bulusheva L~G
	2019 {\em Applied Physics Letters\/} {\bf 114} 091901 ISSN 0003-6951
	(\textit{Preprint}
	\eprint{https://pubs.aip.org/aip/apl/article-pdf/doi/10.1063/1.5080617/13495677/091901\_1\_online.pdf})
	\urlprefix\url{https://doi.org/10.1063/1.5080617}
	
	\bibitem{Mahmood2015}
	Mahmood J, Lee E~K, Jung M, Shin D, Jeon I~Y, Jung S~M, Choi H~J, Seo J~M, Bae
	S~Y, Sohn S~D, Park N, Oh J~H, Shin H~J and Baek J~B 2015 {\em Nature
		Communications\/} {\bf 6} 6486
	\urlprefix\url{https://doi.org/10.1038/ncomms7486}
	
	\bibitem{Zhao2017}
	Zhao Y, Dai Z, Lian C and Meng{,} S 2017 {\em RSC Adv.\/} {\bf 7}(42)
	25803--25810 \urlprefix\url{http://dx.doi.org/10.1039/C7RA03597G}
	
	\bibitem{Omidvar2017}
	Omidvar A 2017 {\em Materials Chemistry and Physics\/} {\bf 202} 258--265 ISSN
	0254-0584
	\urlprefix\url{https://www.sciencedirect.com/science/article/pii/S0254058417307290}
	
	\bibitem{Sousa2022}
	de~Sousa M~S~M, Liu F, Qu F and Chen W 2022 {\em Phys. Rev. B\/} {\bf 105}(1)
	014511 \urlprefix\url{https://link.aps.org/doi/10.1103/PhysRevB.105.014511}
	
	\bibitem{Yang2023}
	Yang J, Ma M, Li L, Zhang Y, Huang W and Dong X 2014 {\em Nanoscale\/} {\bf 6}
	13301--13313 \urlprefix\url{http://dx.doi.org/10.1039/C4NR04584J}
	
	\bibitem{Lin2023}
	Lin Y, Liao Y, Chen Z and Connell J~W 2017 {\em Materials Research Letters\/}
	{\bf 5} 209--234
	\urlprefix\url{https://doi.org/10.1080/21663831.2016.1271047}
	
	\bibitem{Liu2022}
	Liu X, Cho S~M, Lin S, Chen Z, Choi W, Kim Y~M, Yun E, Baek E~H, Ryu D~H and
	Lee H 2022 {\em Matter\/} {\bf 5} 2306--2318
	\urlprefix\url{https://doi.org/10.1016/j.matt.2022.04.033}
	
	\bibitem{Naumis2007}
	Naumis G~G 2007 {\em Phys. Rev. B\/} {\bf 76}(15) 153403
	\urlprefix\url{https://link.aps.org/doi/10.1103/PhysRevB.76.153403}
	
	\bibitem{Xu2019}
	Xu K, Urgel J, Eimre K, Di~Giovannantonio M, Keerthi A, Komber H, Wang S,
	Narita A, Berger R, Ruffieux P, Pignedoli C~A, Liu J, M{\"u}llen K, Fasel R
	and Feng X 2019 {\em Journal of the American Chemical Society\/} {\bf 141}
	7726--7730 \urlprefix\url{https://doi.org/10.1021/jacs.9b03554}
	
	\bibitem{Singh2020}
	Singh D, Shukla V and Ahuja R 2020 {\em Phys. Rev. B\/} {\bf 102}(7) 075444
	\urlprefix\url{https://link.aps.org/doi/10.1103/PhysRevB.102.075444}
	
	\bibitem{Rapjut2023}
	Rajput N~S, Al~Zadjali S, Gutierrez M, Esawi A~M~K and Al~Teneiji M 2021 {\em
		RSC Advances\/} {\bf 11} 27381--27405
	\urlprefix\url{http://dx.doi.org/10.1039/D1RA05157A}
	
	\bibitem{Lokhande2023}
	Lokhande A~C, Qattan I~A, Lokhande C~D and Patole S~P 2020 {\em Journal of
		Materials Chemistry A\/} {\bf 8} 918--977
	\urlprefix\url{http://dx.doi.org/10.1039/C9TA10667G}
	
	\bibitem{Rivera2021}
	Plaza-Rivera C~O, Viggiano R~P, Dornbusch D~A, Wu J~J, Connell J~W and Lin Y
	2021 {\em Frontiers in Energy Research\/} {\bf 9} ISSN 2296-598X
	\urlprefix\url{https://www.frontiersin.org/articles/10.3389/fenrg.2021.703676}
	
	\bibitem{Liu2020}
	Liu T, Zhang L, Cheng B, Hu X and Yu J 2020 {\em Cell Reports Physical
		Science\/} {\bf 1} \urlprefix\url{https://doi.org/10.1016/j.xcrp.2020.100215}
	
	\bibitem{YiLin2017}
	Lin Y, Liao Y, Chen Z and Connell J~W 2017 {\em Materials Research Letters\/}
	{\bf 5} 209--234 (\textit{Preprint}
	\eprint{https://doi.org/10.1080/21663831.2016.1271047})
	\urlprefix\url{https://doi.org/10.1080/21663831.2016.1271047}
	
	\bibitem{PavelBarkov2021}
	Barkov P~V and Glukhova O~E 2021 {\em Nanomaterials\/} {\bf 11} 1074 ISSN
	2079-4991 \urlprefix\url{http://dx.doi.org/10.3390/nano11051074}
	
	\bibitem{Fischbein2008}
	Fischbein M~D and Drndić M 2008 {\em Applied Physics Letters\/} {\bf 93}
	113107 ISSN 0003-6951 (\textit{Preprint}
	\eprint{https://pubs.aip.org/aip/apl/article-pdf/doi/10.1063/1.2980518/14401035/113107\_1\_online.pdf})
	\urlprefix\url{https://doi.org/10.1063/1.2980518}
	
	\bibitem{Khan2021}
	Khan K, Liu T, Arif M, Yan X, Hossain M~D, Rehman F, Zhou S, Yang J, Sun C, Bae
	S~H, Kim J, Amine K, Pan X and Luo Z 2021 {\em Advanced Energy Materials\/}
	{\bf 11} 2101619 (\textit{Preprint}
	\eprint{https://onlinelibrary-wiley-com.pbidi.unam.mx:2443/doi/pdf/10.1002/aenm.202101619})
	\urlprefix\url{https://onlinelibrary-wiley-com.pbidi.unam.mx:2443/doi/abs/10.1002/aenm.202101619}
	
	\bibitem{KAZEMIZADEH2018}
	Kazemizadeh F and Malekfar R 2018 {\em Physica B: Condensed Matter\/} {\bf 530}
	236--241 ISSN 0921-4526
	\urlprefix\url{https://www.sciencedirect.com/science/article/pii/S092145261730947X}
	
	\bibitem{Wang2019}
	Wang F, Mei X, Wang K, Dong X, Gao M, Zhai Z, Lv J, Zhu C, Duan W and Wang W
	2019 {\em Journal of Materials Science\/} {\bf 54} 5658--5670
	\urlprefix\url{https://doi.org/10.1007/s10853-018-03247-0}
	
	\bibitem{Lin2014}
	Lin J, Peng Z, Liu Y, Ruiz-Zepeda F, Ye R, Samuel E~L~G, Yacaman M~J, Yakobson
	B~I and Tour J~M 2014 {\em Nature Communications\/} {\bf 5} 5714
	\urlprefix\url{https://doi.org/10.1038/ncomms6714}
	
	\bibitem{KUMAR2022}
	Kumar R, {Pérez del Pino} A, Sahoo S, Singh R~K, Tan W~K, Kar K~K, Matsuda A
	and Joanni E 2022 {\em Progress in Energy and Combustion Science\/} {\bf 91}
	100981 ISSN 0360-1285
	\urlprefix\url{https://www.sciencedirect.com/science/article/pii/S0360128521000794}
	
	\bibitem{Joshi2022}
	Joshi P, Shukla S, Gupta S, Riley P~R, Narayan J and Narayan R 2022 {\em ACS
		Applied Materials \& Interfaces\/} {\bf 14} 37149--37160
	\urlprefix\url{https://doi.org/10.1021/acsami.2c09096}
	
	\bibitem{HaruyamaJunji2013}
	Haruyama J 2013 {\em Electronics\/} {\bf 2} 368–386 ISSN 2079-9292
	\urlprefix\url{http://dx.doi.org/10.3390/electronics2040368}
	
	\bibitem{LinYi2022}
	Lin Y, Plaza-Rivera C~O, Hu L and Connell J~W 2022 {\em Accounts of Chemical
		Research\/} {\bf 55} 3020--3031 pMID: 36173244 (\textit{Preprint}
	\eprint{https://doi.org/10.1021/acs.accounts.2c00457})
	\urlprefix\url{https://doi.org/10.1021/acs.accounts.2c00457}
	
	\bibitem{Zhang2019}
	Zhang Y, Wan Q and Yang N 2019 {\em Small\/} {\bf 15} 1903780
	(\textit{Preprint}
	\eprint{https://onlinelibrary-wiley-com.pbidi.unam.mx:2443/doi/pdf/10.1002/smll.201903780})
	\urlprefix\url{https://onlinelibrary-wiley-com.pbidi.unam.mx:2443/doi/abs/10.1002/smll.201903780}
	
	\bibitem{White2020}
	White D~L, Lystrom L, He X, Burkert S~C, Kilin D~S, Kilina S and Star A 2020
	{\em ACS Applied Materials \& Interfaces\/} {\bf 12} 36513--36522
	\urlprefix\url{https://doi.org/10.1021/acsami.0c09394}
	
	\bibitem{Zhang2016}
	Zhang J, Song H, Zeng D, Wang H, Qin Z, Xu K, Pang A and Xie C 2016 {\em
		Scientific Reports\/} {\bf 6} 32310
	\urlprefix\url{https://doi.org/10.1038/srep32310}
	
	\bibitem{Bai2010}
	Bai J, Zhong X, Jiang S, Huang Y and Duan X 2010 {\em Nature Nanotechnology\/}
	{\bf 5} 190--194 \urlprefix\url{https://doi.org/10.1038/nnano.2010.8}
	
	\bibitem{pybinding}
	Moldovan D, An\l{d}elkovi{\'{c}} M and Peeters F 2020 {pybinding v0.9.5: a
		Python package for tight- binding calculations} {This work was supported by
		the Flemish Science Foundation (FWO-Vl) and the Methusalem Funding of the
		Flemish Government.} \urlprefix\url{https://doi.org/10.5281/zenodo.4010216}
	
	\bibitem{pyqula}
	Lado J~L 2021 {G}it{H}ub - joselado/pyqula: {P}ython library to compute
	different properties of quantum tight binding models in a lattice.
	\url{https://github.com/joselado/pyqula}
	
	\bibitem{Tbplas2023}
	Li Y, Zhan Z, Kuang X, Li Y and Yuan S 2023 {\em Computer Physics
		Communications\/} {\bf 285} 108632 ISSN 0010-4655
	\urlprefix\url{https://www.sciencedirect.com/science/article/pii/S0010465522003514}
	
	\bibitem{Koh2023}
	Koh K~H, Bagherzadeh~Mostaghimi A~H, Chang Q, Kim Y~J, Siahrostami S, Han T~H
	and Chen Z 2023 {\em EcoMat\/} {\bf 5} e12266 (\textit{Preprint}
	\eprint{https://onlinelibrary-wiley-com.pbidi.unam.mx:2443/doi/pdf/10.1002/eom2.12266})
	\urlprefix\url{https://onlinelibrary-wiley-com.pbidi.unam.mx:2443/doi/abs/10.1002/eom2.12266}
	
	\bibitem{Naumis1994}
	Naumis G~G, Barrio R~A and Wang C 1994 {\em Phys. Rev. B\/} {\bf 50}(14)
	9834--9842 \urlprefix\url{https://link.aps.org/doi/10.1103/PhysRevB.50.9834}
	
	\bibitem{Naumis2002}
	Naumis G~G, Wang C and Barrio R~A 2002 {\em Phys. Rev. B\/} {\bf 65}(13) 134203
	\urlprefix\url{https://link.aps.org/doi/10.1103/PhysRevB.65.134203}
	
	\bibitem{Bell1970}
	Bell R~J and Dean P 1970 {\em Discuss. Faraday Soc.\/} {\bf 50}(0) 55--61
	\urlprefix\url{http://dx.doi.org/10.1039/DF9705000055}
	
	\bibitem{Edwards1972}
	Edwards J~T and Thouless D~J 1972 {\em Journal of Physics C: Solid State
		Physics\/} {\bf 5} 807
	\urlprefix\url{https://dx.doi.org/10.1088/0022-3719/5/8/007}
	
	\bibitem{Shukla2018}
	Shukla P 2018 {\em Phys. Rev. B\/} {\bf 98}(5) 054206
	\urlprefix\url{https://link.aps.org/doi/10.1103/PhysRevB.98.054206}
	
	\bibitem{Wegner1980}
	Wegner F 1980 {\em Zeitschrift f{\"u}r Physik B Condensed Matter\/} {\bf 36}
	209--214 \urlprefix\url{https://doi.org/10.1007/BF01325284}
	
	\bibitem{Jun2023}
	Bae J~H, Sedrakyan T and Maiti S 2023 {\em SciPost Phys.\/} {\bf 15} 139
	\urlprefix\url{https://scipost.org/10.21468/SciPostPhys.15.4.139}
	
	\bibitem{Gerardo2014}
	Naumis G~G and Roman-Taboada P 2014 {\em Phys. Rev. B\/} {\bf 89}(24) 241404
	\urlprefix\url{https://link.aps.org/doi/10.1103/PhysRevB.89.241404}
	
	\bibitem{BARRIOSVARGAS201323}
	Barrios-Vargas J and Naumis G~G 2013 {\em Solid State Communications\/} {\bf
		162} 23--27 ISSN 0038-1098
	\urlprefix\url{https://www.sciencedirect.com/science/article/pii/S0038109813001221}
	
	\bibitem{Wang2016}
	Wang E, Lu X, Ding S, Yao W, Yan M, Wan G, Deng K, Wang S, Chen G, Ma L, Jung
	J, Fedorov A~V, Zhang Y, Zhang G and Zhou S 2016 {\em Nature Physics\/} {\bf
		12} 1111--1115 ISSN 1745-2481
	\urlprefix\url{https://doi.org/10.1038/nphys3856}
	
	\bibitem{Kou2014}
	Kou L, Hu F, Yan B, Frauenheim T and Chen C 2014 {\em Nanoscale\/} {\bf 6}(13)
	7474--7479 \urlprefix\url{http://dx.doi.org/10.1039/C4NR01102C}
	
	\bibitem{NISHIDATE2023122196}
	Nishidate K, Matsukawa M and Hasegawa M 2023 {\em Surface Science\/} {\bf 728}
	122196 ISSN 0039-6028
	\urlprefix\url{https://www.sciencedirect.com/science/article/pii/S0039602822001819}
	
	\bibitem{AGRAWAL2013102}
	Agrawal B and Agrawal S 2013 {\em Physica E: Low-dimensional Systems and
		Nanostructures\/} {\bf 50} 102--107 ISSN 1386-9477
	\urlprefix\url{https://www.sciencedirect.com/science/article/pii/S1386947713000684}
	
	\bibitem{Malterre_2011}
	Malterre D, Kierren B, Fagot-Revurat Y, Didiot C, de~Abajo F~J~G, Schiller F,
	Cordón J and Ortega J~E 2011 {\em New Journal of Physics\/} {\bf 13} 013026
	\urlprefix\url{https://dx.doi.org/10.1088/1367-2630/13/1/013026}
	
	\bibitem{Zhou2007}
	Zhou S~Y, Gweon G~H, Fedorov A~V, First P~N, de~Heer W~A, Lee D~H, Guinea F,
	Castro~Neto A~H and Lanzara A 2007 {\em Nature Materials\/} {\bf 6} 770--775
	ISSN 1476-4660 \urlprefix\url{https://doi.org/10.1038/nmat2003}
	
	\bibitem{Jia2016}
	Jia T~T, Fan X~Y, Zheng M~M and Chen G 2016 {\em Scientific Reports\/} {\bf 6}
	20971 ISSN 2045-2322 \urlprefix\url{https://doi.org/10.1038/srep20971}
	
	\bibitem{YE201460}
	Ye M, Quhe R, Zheng J, Ni Z, Wang Y, Yuan Y, Tse G, Shi J, Gao Z and Lu J 2014
	{\em Physica E: Low-dimensional Systems and Nanostructures\/} {\bf 59} 60--65
	ISSN 1386-9477
	\urlprefix\url{https://www.sciencedirect.com/science/article/pii/S1386947713004530}
	
	\bibitem{Mojarro2020}
	Mojarro M~A, Ibarra-Sierra V~G, Sandoval-Santana J~C, Carrillo-Bastos R and
	Naumis G~G 2020 {\em Phys. Rev. B\/} {\bf 101}(16) 165305
	\urlprefix\url{https://link.aps.org/doi/10.1103/PhysRevB.101.165305}
	
\end{thebibliography}
\providecommand{\newblock}{}


\end{document}